\documentclass[reprint,amsmath,amssymb,aps]{revtex4-2}
\usepackage{float}
\usepackage{appendix}
\usepackage{lipsum}
\usepackage[colorlinks, citecolor=blue,urlcolor=blue,linkcolor=blue,runcolor=filecolor]{hyperref}
\usepackage{url}
\usepackage{amsmath}
\usepackage{subfigure}
\allowdisplaybreaks[4]
\usepackage{graphicx}
\usepackage{dcolumn}
\usepackage{bm}
\usepackage{amsfonts}
\usepackage{amssymb}
\usepackage{array}
\usepackage{graphics}
\usepackage{hyperref}
\usepackage{upgreek}
\begin{document}

\title{Probing new light scalars with the lepton anomalous magnetic moment and the weak equivalence principle violation }
	
\author{Xitong Mei\textsuperscript{1,2,}}
\author{Dongfeng Gao\textsuperscript{1,3,}}
\altaffiliation{Email: dfgao@wipm.ac.cn}
\author{Wei Zhao\textsuperscript{4}}
\author{Jin Wang\textsuperscript{1,3,5,}}
\altaffiliation{Email: wangjin@wipm.ac.cn}
\author{Mingsheng Zhan\textsuperscript{1,3,5,}}
\altaffiliation{Email: mszhan@wipm.ac.cn}
	
\vskip 0.5cm
\affiliation{1 State Key Laboratory of Magnetic Resonance and Atomic and Molecular Physics, Wuhan Institute of Physics and Mathematics, Innovation Academy for Precision Measurement Science and Technology, Chinese Academy of Sciences, Wuhan 430071, China\\
		2 School of Physical Sciences, University of Chinese Academy of Sciences, Beijing 100049, China\\
		3 Hefei National Laboratory, Hefei 230088, China\\
		4 Shandong University of Aeronautics, Binzhou 256600, China\\
		5 Wuhan Institute of Quantum Technology, Wuhan 430206, China}

\begin{abstract}
A new scalar particle with generic couplings to the standard-model particles is a possible source for the lepton anomalous magnetic moment and the violation of the weak equivalence principle. Here, one-loop contributions to the lepton anomalous magnetic moment, involving the scalar-photon and scalar-lepton couplings, are calculated. Then, employing the recent experimental results of the electron anomalous magnetic moment, the muon anomalous magnetic moment, and the MICROSCOPE mission, we find the improved constraints on scalar-lepton and scalar-photon couplings: $|\lambda_e|\leq 6.0 \times 10^{-6}$, $|\lambda_\mu|\leq 3.5\times 10^{-4}$, and $|\lambda_\gamma|\leq 4.5 \times 10^{-13}$ ${\rm eV^{-1}}$ for scalar mass below $10^4$ eV. We find that the naive scaling relationship between the scalar-muon coupling and the scalar-electron coupling is favored by three experimental results. Furthermore, the minimal standard-model extension by one scalar is also favored by all three experiments, and the model parameter is constrained best to $|\mathcal{A}|\leq 1.7 \times 10^{-11}$ eV for $m_{\phi}< 10^{-13}$ eV. 
\end{abstract}


\maketitle

\section{Introduction}\label{sec:1}
Many theories beyond the standard model (SM) of particle physics suggest the existence of additional spin-0 bosons, which could be motivated by various issues, e. g. axions in the strong CP problem \cite{PhysRevLett.38.1440, PhysRevLett.40.223, PhysRevLett.40.279}, axion-like-particle candidates in dark matter detection \cite{essig2013}, extensions in the scalar sector of the SM \cite{Patt2006fw,PhysRevD.75.037701,PhysRevD.77.035005,PhysRevD.82.043533}, moduli in string theory \cite{Conlon_2006, Svrcek_2006, PhysRevD.81.123530}, and dilatons in gravitational physics \cite{PhysRev.124.925, PhysRevD.82.084033}. 
Such a broad class of particles can be involved in various kinds of couplings to the SM particles, which can be either fundamental or effective, depending on the specific mechanism associated with the sign of new physics. The search for such particles, and consequently constraining the mass and coupling parameters, can be carried out by various experimental methods \cite{battaglieri2017,PhysRevD.110.030001}. 

In particular, the lepton anomalous magnetic moment, with the rapid improvement of experimental measurements, has been used as a stringent probe for the hypothetical spin-0 bosons. 
The SM prediction of the lepton anomalous magnetic moment can be derived as a function of the fine structure constant $\alpha$, and has been calculated in many studies (for example, see Refs. \cite{PhysRevLett.109.111807,atoms7010028,volkov2024} for the electron, and Refs. \cite{PhysRevLett.109.111808,AOYAMA20201,colangelo2022prospects} for the muon). 
The fine structure constant can be best measured by atomic recoil experiments \cite{Parker2018, Morel2020}. Suppose that the discrepancy between the SM prediction and the measurement for the lepton anomalous magnetic moments is caused by hypothetical spin-0 bosons. Then, the discrepancy could be used to constrain their masses and coupling parameters (for example, see Refs. \cite{PhysRevD.63.091301,PhysRevD.93.035006, PhysRevD.94.115033}).

For the electron, the SM prediction of the anomalous magnetic moment is $a_e^{\mathrm{SM}}=1159652180.252(95)\times{10}^{-12}$, employing the experimental result $\alpha^{\,-1}=137.035999206(11)$ obtained in \cite{Morel2020}.
The electron anomalous magnetic moment can also be directly measured with one-electron quantum cyclotrons, where the best result is $a_e^{\mathrm{Meas}}=1159652180.59(13)\times{10}^{-12}$ \cite{Fan2023}.
Thus, the discrepancy between the SM prediction and the measurement is
\begin{equation}
\label{daelectron} 
\delta a_e^{\rm EXP}=a_e^{\mathrm{Meas}}-a_e^{\mathrm{SM}}=0.34(16)\times{10}^{-12},
\end{equation}
which is about 2.1$\sigma$. In other words, there seems to be a sign of new physics at the 2.1$\sigma$ level, although the sign is not so strong. 

For the muon, the issue of the discrepancy between the SM prediction and the experimental measurement is quite tricky. In 2020, the Muon $g-2$ Theory Initiative published the first White Paper (WP2020), yielding the best SM prediction at the time, $a_{\mu}^{\mathrm{SM}}(2020)=116591810(43)\times 10^{-11}$ \cite{AOYAMA20201}. In 2023, by analysing the data collected in 2019 and 2020, the Fermilab Muon $g-2$ collaboration announced the best experimental world average value at the time, $a_{\mu}^{\mathrm{Meas}}(2023)=116592059(22)\times 10^{-11} $ \cite{PhysRevLett.131.161802}. Combining these two values would give a discrepancy, $\delta a_{\mu}^{\rm EXP}(2023)=a_{\mu}^{\mathrm{Meas}}(2023)-a_{\mu}^{\mathrm{SM}}(2020)=2.49(48)\times 10^{-9}$, which is about 5.2$\sigma$. The sign of new physics was quite strong. However, mainly due to the improved estimate of the leading-order hadronic-vacuum-polarization contribution since WP2020, the Muon $g-2$ Theory Initiative published an updated White Paper (WP2025), yielding the updated SM prediction, $a_{\mu}^{\mathrm{SM}}(2025)=116592033(62)\times 10^{-11}$ \cite{wp25}. In June 2025, the Fermilab Muon $g-2$ collaboration finished analysing the data taken from 2020 to 2023 and announced the new experimental world average, $a_{\mu}^{\mathrm{Meas}}(2025)=1165920715(124)\times 10^{-12} $ \cite{muon2025}. With these two latest values, one could get a value 
\begin{equation}
	\label{damuon} \delta a_{\mu}^{\rm EXP}(2025)=3.9(6.4)\times 10^{-10} ,
\end{equation}
which will be used in this work. Apparently, there is no sign of new physics even at the 0.6$\sigma$ level. 

On the other hand, high precision tests of the weak equivalence principle (WEP) violation can also be used to probe hypothetical spin-0 particles \cite{PhysRev.124.925, PhysRevD.82.084033, PhysRevLett.120.141101}. The MICROSCOPE mission \cite{PhysRevLett.129.121102,Touboul2022} has achieved the highest precision of the WEP test, where the Eötvös parameter $\eta$ is measured to be
\begin{equation} \label{etaresult} 
 \eta(\mathrm{Pt, Ti})^{\rm EXP}=-1.5(2.7)\times{10}^{-15}. 
\end{equation} 
Although there is no sign of new physics even at the -0.6$\sigma$ level, the result (\ref{etaresult}) can still be used to set useful constraints on new physics models. Suppose that the WEP violation is caused by new light scalar particles through their couplings to the SM particles. Then, the MICROSCOPE mission can set new constraints on such scalar particles \cite{PhysRevLett.120.141101, Berge_2023}. 

In this work, we investigate the possibility that a new light scalar can account for the discrepancy (\ref{daelectron}) for the electron, the discrepancy (\ref{damuon}) for the muon, and the WEP violation (\ref{etaresult}). Here, we concentrate on the scalar coupling to the photon and its Yukawa couplings to leptons. In that context, we restrict ourself to the scalar mass range below $10^4$ eV, where the light scalar is regarded as a dark matter candidate and gets more and more attentions from various experiments \cite{battaglieri2017,Beloy2021,Vermeulen2021}. The advantage of combining these three experiments together lies in the fact that these three experiments cover all the four fundamental interactions in nature. We get new constraints on the new scalar particle, which could not be obtained by using either the lepton anomalous magnetic moment or the WEP violation individually. 

This paper is organized as follows. In Sec.~\ref{sec:2}, we write down the Lagrangian with linear couplings to the photon and leptons. Then, the results of the one-loop contribution to lepton anomalous magnetic moments and the scalar contribution to the Eötvös parameter are summarized. Detailed calculations are given in Appendixes \ref{caloneloop} and \ref{caleta}. In Sec.~\ref{sec:4}, we discuss the case where the new scalar couples to both the photon and leptons simultaneously. Improved constraints on scalar-photon coupling and scalar-lepton couplings are found. The naive scaling relationship between scalar-muon coupling and scalar-electron coupling is favored by three experimental results. In Sec.~\ref{minimalSME}, the minimal SM extension model in the scalar sector is discussed. Finally, conclusion and discussion are given in Sec. \ref{sec:5}.

\section{The scalar model $\&$ calculation of $\delta a_l$ and $\eta$}\label{sec:2}

To be specific, let us work on a linear coupling model, where linear couplings between the new scalar $\phi$, the photon and the SM leptons are assumed. Following Ref. \cite{PhysRevD.82.084033}, the interaction terms can be formally written as 

\begin{eqnarray}\label{scalarmodel}
\mathcal{L}'_{int}=&&\phi\left[\lambda_{\gamma} F_{\mu\nu}F^{\mu\nu}+\sum_{l=e,\mu}\lambda_l\bar{\psi}_l\psi_l\right],
\end{eqnarray}
where $\psi_l$ stands for the lepton fields for $l=e, \mu$. $\lambda_{\gamma}$ and $\lambda_l$ denote the coupling to the $U(1)$ photon, and the dimensionless Yukawa coupling to leptons, respectively.

\subsection{One-Loop Contribution to $\delta a_l$} \label{sec:2.1}

Suppose that the discrepancy between the SM prediction and the measurement for the lepton anomalous magnetic moments, $\delta a_l$ ($l=e$ or $\mu$), is caused by the new scalar $\phi$. Since the  Yukawa coupling $\lambda_l$ is assumed to be small, it is enough to consider the contribution from one-loop Feynman diagrams to $\delta a_l$.

\begin{figure}[h]
\centering	
\includegraphics[width=0.2\textwidth]{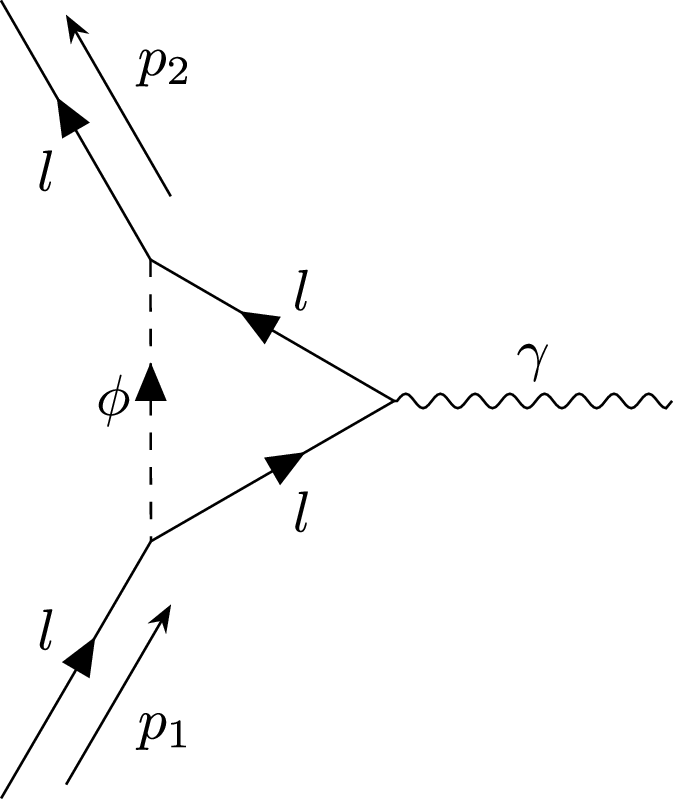}
\caption{\label{fig3a} The Scalar-Lepton-Lepton loop diagram}
\end{figure}

\begin{figure}[h]
	\centering
	\includegraphics[width=0.4\textwidth]{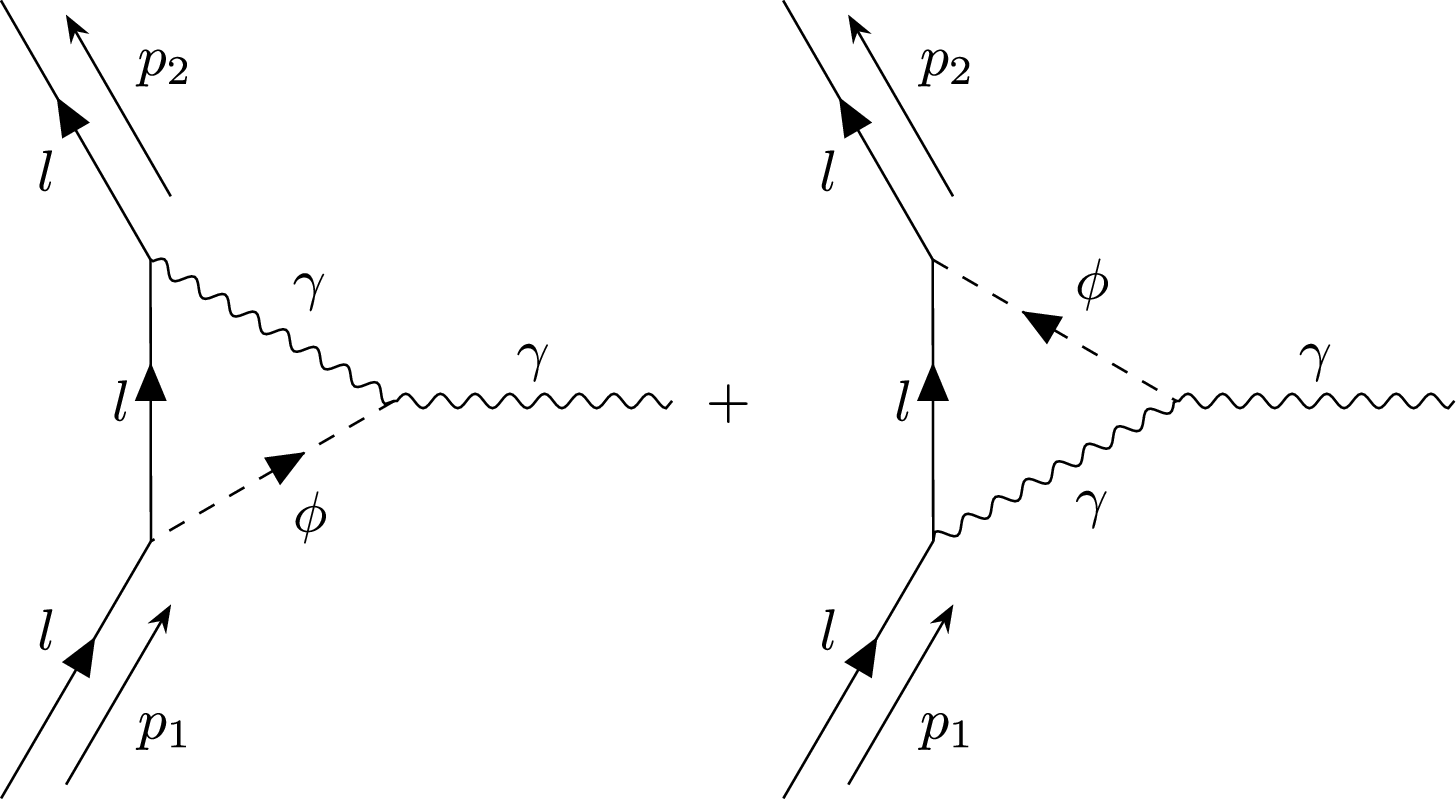}
	\caption{\label{fig5a} The Scalar-Lepton-Photon loop diagrams}
\end{figure}

At one-loop level, there exist two types of Feynman diagrams. One is called the Scalar-Lepton-Lepton loop diagram (Fig. \ref{fig3a}) and the other is called the Scalar-Lepton-Photon loop diagram (Fig. \ref{fig5a}). Using the Passarino-Veltman Renormalization \cite{PASSARINO1979151,THOOFT1979365}, we can calculate them. The details of the calculation are given in Appendix \ref{caloneloop}. Here we summarize the main results. 

For the Scalar-Lepton-Lepton loop diagram, its contribution to $\delta a_l$ is calculated to be
\begin{equation}
\delta a_{l1}(m_{\phi})={\lambda_l}^2a_{sll}(r_l)\, ,
\end{equation}
with
\begin{widetext}
	\begin{eqnarray}
		a_{sll}(r_l)=
		\begin{cases}
			\frac{-2{r_l}^2-3{r_l}^2\log({r_l}^2)+{r_l}^4\log({r_l}^2)-2\sqrt{4{r_l}^2-{r_l}^4}{r_l}^2\cos^{-1}(\frac{r_l}{2})+2\sqrt{4{r_l}^2-{r_l}^4}\cos^{-1}(\frac{r_l}{2})+3}{16\pi^2} & \text{if } r_l \leq 2,\\
			
			\frac{-2{r_l}^2-3{r_l}^2\log({r_l}^2)+{r_l}^4\log({r_l}^2)-2\sqrt{{r_l}^4-4{r_l}^2}{r_l}^2\cosh^{-1}(\frac{r_l}{2})+2\sqrt{{r_l}^4-4{r_l}^2}\cosh^{-1}(\frac{r_l}{2})+3}{16\pi^2} & \text{if } r_l \geq 2.
		\end{cases}
	\end{eqnarray}
\end{widetext}
Here $r_l\equiv m_{\phi}/m_l$, where $m_{\phi}$ is the mass of $\phi$, and $m_l$ is the mass of leptons. In Ref. \cite{PhysRevD.93.035006}, the authors calculated the same diagram. Our result is consistent with theirs.

Similarly, the contribution from the Scalar-Lepton-Photon loop diagrams is found to be
\begin{equation}
\delta a_{l2}(m_{\phi})=\lambda_l\lambda_{\gamma} m_{l}b_{sl\gamma}(r_l)\, ,
\end{equation}
with
\begin{widetext}
\begin{eqnarray}
b_{sl\gamma}(r_l)=
\begin{cases}
\frac{-2{r_l}^2-6{r_l}^2\log({r_l}^2)+{r_l}^4\log({r_l}^2)-2\sqrt{4{r_l}^2-{r_l}^4}{r_l}^2\cos^{-1}(\frac{r_l}{2})+8\sqrt{4{r_l}^2-{r_l}^4}\cos^{-1}(\frac{r_l}{2})+18}{48\pi^2} & \text{if } r_l \leq 2,\\

\frac{-2{r_l}^2-6{r_l}^2\log({r_l}^2)+{r_l}^4\log({r_l}^2)-2\sqrt{{r_l}^4-4{r_l}^2}{r_l}^2\cosh^{-1}(\frac{r_l}{2})+8\sqrt{{r_l}^4-4{r_l}^2}\cosh^{-1}(\frac{r_l}{2})+18}{48\pi^2} & \text{if } r_l \geq 2.
\end{cases}
\end{eqnarray}
\end{widetext}

\begin{figure}[htbp]
	\centering
\includegraphics[height=0.22\textheight, width=0.45\textwidth]{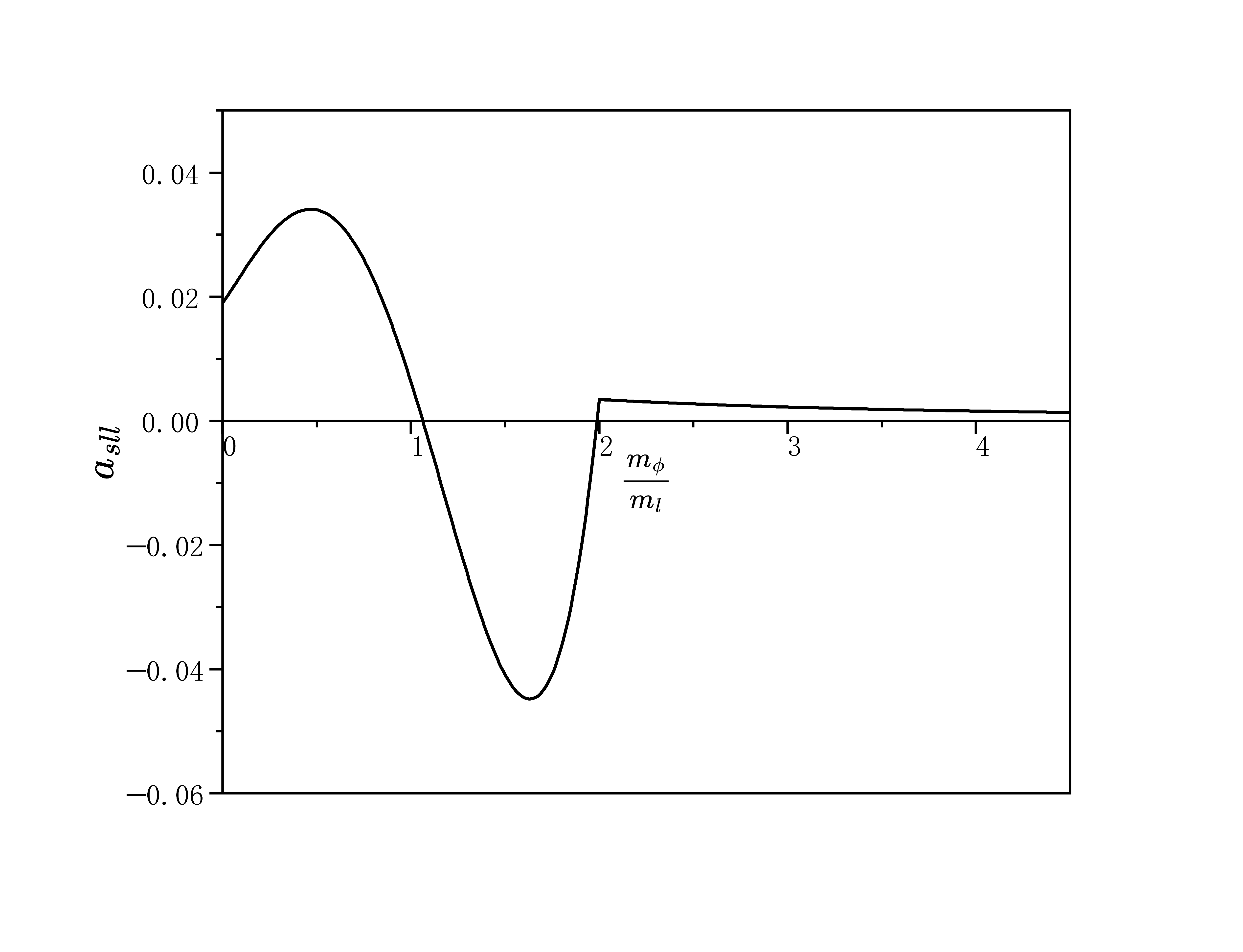}
\includegraphics[height=0.22\textheight, width=0.45\textwidth]{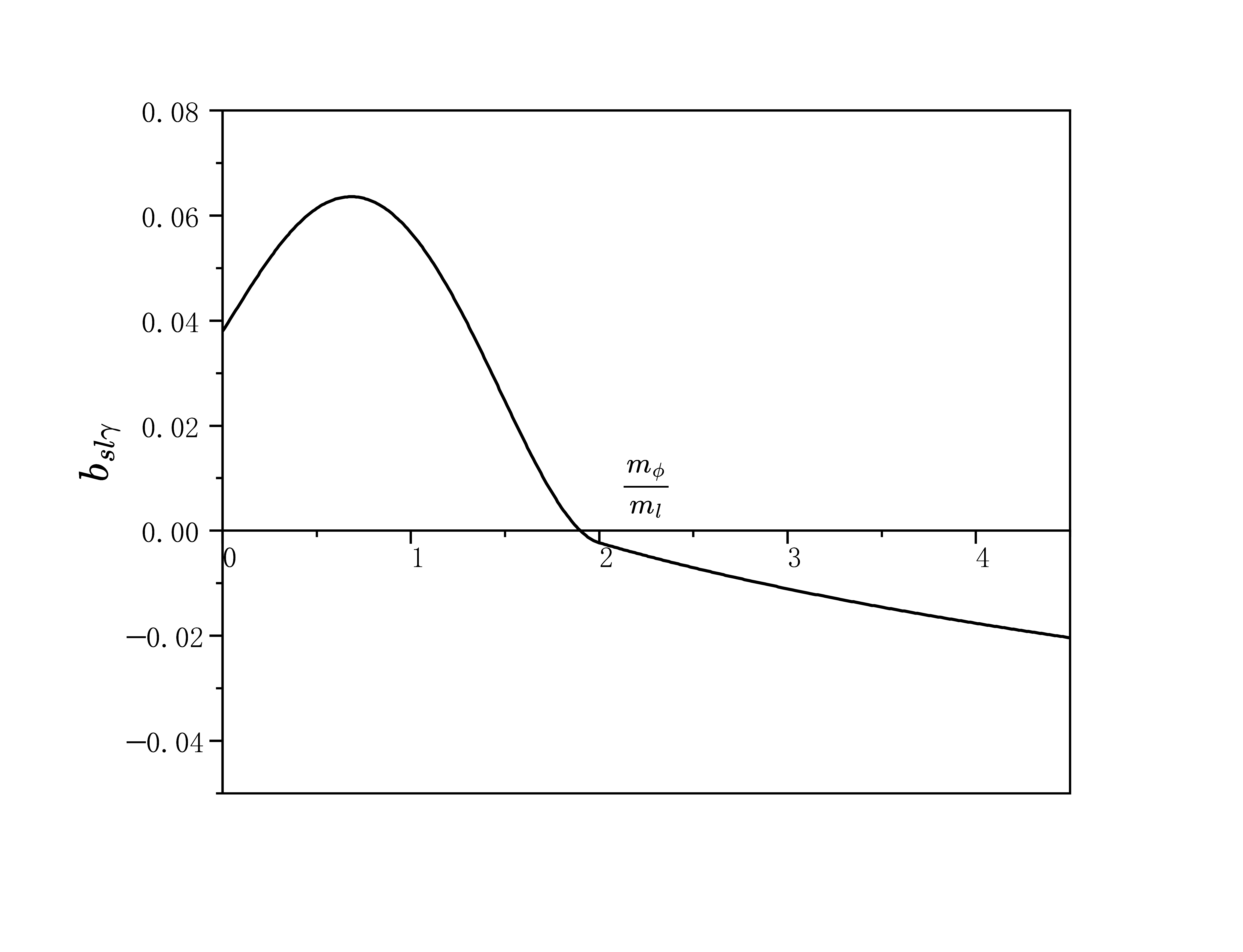}
\caption{The behaviors of $a_{sll}(r_l)$ and $b_{sl\gamma}(r_l)$.}
	\label{fig46}
\end{figure}

Then, summing up $\delta a_{l1}$ and $\delta a_{l2}$, the total one-loop contribution of the scalar field to $\delta a_{l}$ is found to be
\begin{equation}\label{deltaa}
	\delta a_{l}=\delta a_{l1}+\delta a_{l2}={\lambda_l}^2 a_{sll}(r_l)+\lambda_l \lambda_{\gamma} m_{l} b_{sl\gamma}(r_l)\, .
\end{equation}
Note that $\lambda_l$ appears in both terms, and $\lambda_{\gamma}$ appears only in the second term.  

It is also worth noticing the behaviors of $a_{sll}(r_l)$ and $b_{sl\gamma}(r_l)$ as functions of $r_l$. As shown in Fig. \ref{fig46}, both $a_{sll}(r_l)$ and $b_{sl\gamma}(r_l)$ are positive for $m_{\phi}< m_e$, which is the mass range we focus on in this work.

\subsection{Contribution of the scalar $\phi$ to $\eta$} 

The $\phi$ contribution to the Eötvös parameter has been calculated in Refs. \cite{PhysRevD.82.084033,Zhao2022}. The summary of the calculation is given in Appendix \ref{caleta}. Here, we quote the results as follows. 

For two test bodies freely falling towards the Earth, the Eötvös parameter $\eta$ is found to be
\begin{eqnarray}\label{etaequation}
	\eta=&&\left(1+\frac{R_E}{\Lambda_{\phi}}\right)I\left(\frac{R_E}{\Lambda_{\phi}}\right)(\zeta'_{A}-\zeta'_{B})\zeta'_E e^{-R_E/\Lambda_{\phi}}\, , \nonumber\\
	&&	I(x)\equiv\frac{3(x\cosh{(x)}-\sinh{(x)})}{x^3}\, ,
\end{eqnarray}
where $R_E$ is the radius of the Earth, $\Lambda_{\phi}\equiv\hbar/m_{\phi}$ is the Compton wavelength of the scalar $\phi$. $\zeta'_{A, B}$ is the so-called scalar-charge for a mass, which is given in Eq. (\ref{atomcharge}). $\zeta'_{\rm E}$ is the scalar-charge for the Earth, which is 
\begin{equation}\label{alphaEarth}
\zeta'_{\rm E}=-1.808\times10^{18}\lambda_e +2.319\times10^{25}\lambda_{\gamma}\cdot{\rm eV}.
\end{equation}

According to Refs. \cite{PhysRevLett.120.141101,PhysRevLett.129.121102,Touboul2022}, the MICROSCOPE mission result (\ref{etaresult}) was achieved for a pair of test masses, which have different
compositions [PtRh(90/10) and TiAlV(90/6/4) alloys]. The scalar-charges for them are
\begin{align}
\label{alphaPt} &\zeta'_{\rm Pt}=-1.496\times10^{18}\lambda_e+5.766\times10^{25}\lambda_{\gamma}\cdot{\rm eV},\,\,\,\,\\
\label{alphaTi} 
&\zeta'_{\rm Ti}=-1.707\times10^{18}\lambda_e +3.101\times10^{25}\lambda_{\gamma}\cdot{\rm eV}.
\end{align}
With Eqs. (\ref{etaequation}-\ref{alphaTi}), one can write down the Eötvös parameter $\eta(\mathrm{Pt, Ti})$ for the MICROSCOPE mission
\begin{align}\label{etaMICROSCOPE1}
&\eta(\mathrm{Pt,Ti})=\left(1+\frac{R_E}{\Lambda_{\phi}}\right)I\left(\frac{R_E}{\Lambda_{\phi}}\right)\left(-3.824\times10^{35}\lambda_e^2\right.\nonumber\\
&\left. +6.179\times10^{50}\lambda_{\gamma}^2\cdot{\rm eV}^2
-4.328\times10^{43}\lambda_e\lambda_{\gamma}\cdot{\rm eV}\right)\nonumber\\
&\times e^{-\frac{R_E}{\Lambda_{\phi}}}
\end{align}
Note that, in the following, constraints on coupling parameters are set at the 2.1$\sigma$ level, which is the maximal confidential level of new physics for all three experimental results.

\section{Full Constraints on $\lambda_e$, $\lambda_{\mu}$ and $\lambda_{\gamma}$ with all three experimental results}\label{sec:4}

\subsection{Constraints on individual parameters} 

With three independent experimental results, we can fully constrain all three coupling parameters: $\lambda_e$, $\lambda_\mu$ and $\lambda_\gamma$. The procedure is as follows. First, by inserting results (\ref{daelectron}) and (\ref{etaresult}) into Eqs. (\ref{deltaa}) and (\ref{etaMICROSCOPE1}), we can solve them and find out the constraints on $\lambda_e$ and $\lambda_\gamma$. Then, inserting the result (\ref{damuon}) and the solution to $\lambda_\gamma$ into Eq. (\ref{deltaa}) for the muon, we can find the constraint on $\lambda_\mu$.

\begin{figure}[htbp]
		\includegraphics[width=0.45\textwidth]{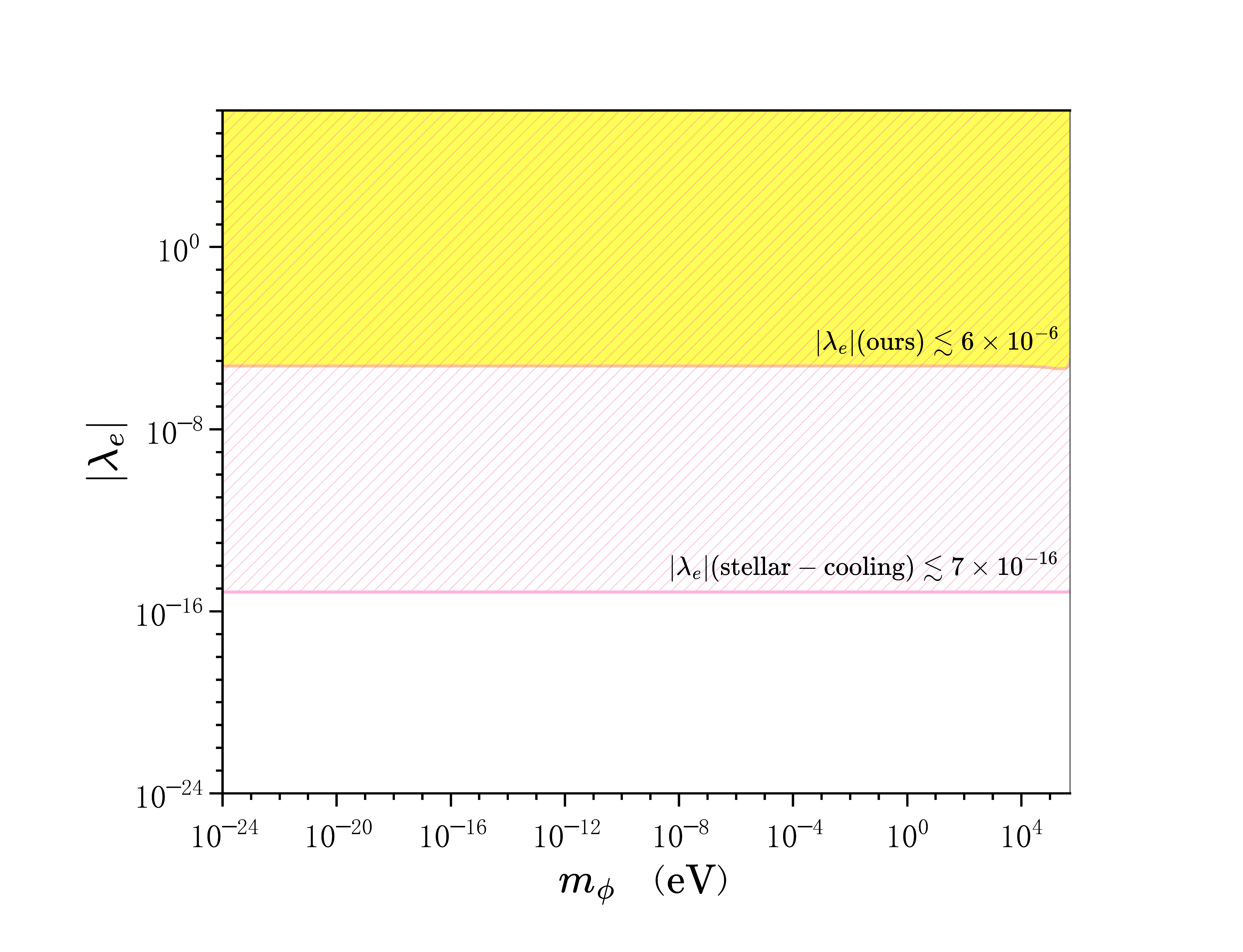}
		\includegraphics[width=0.45\textwidth]{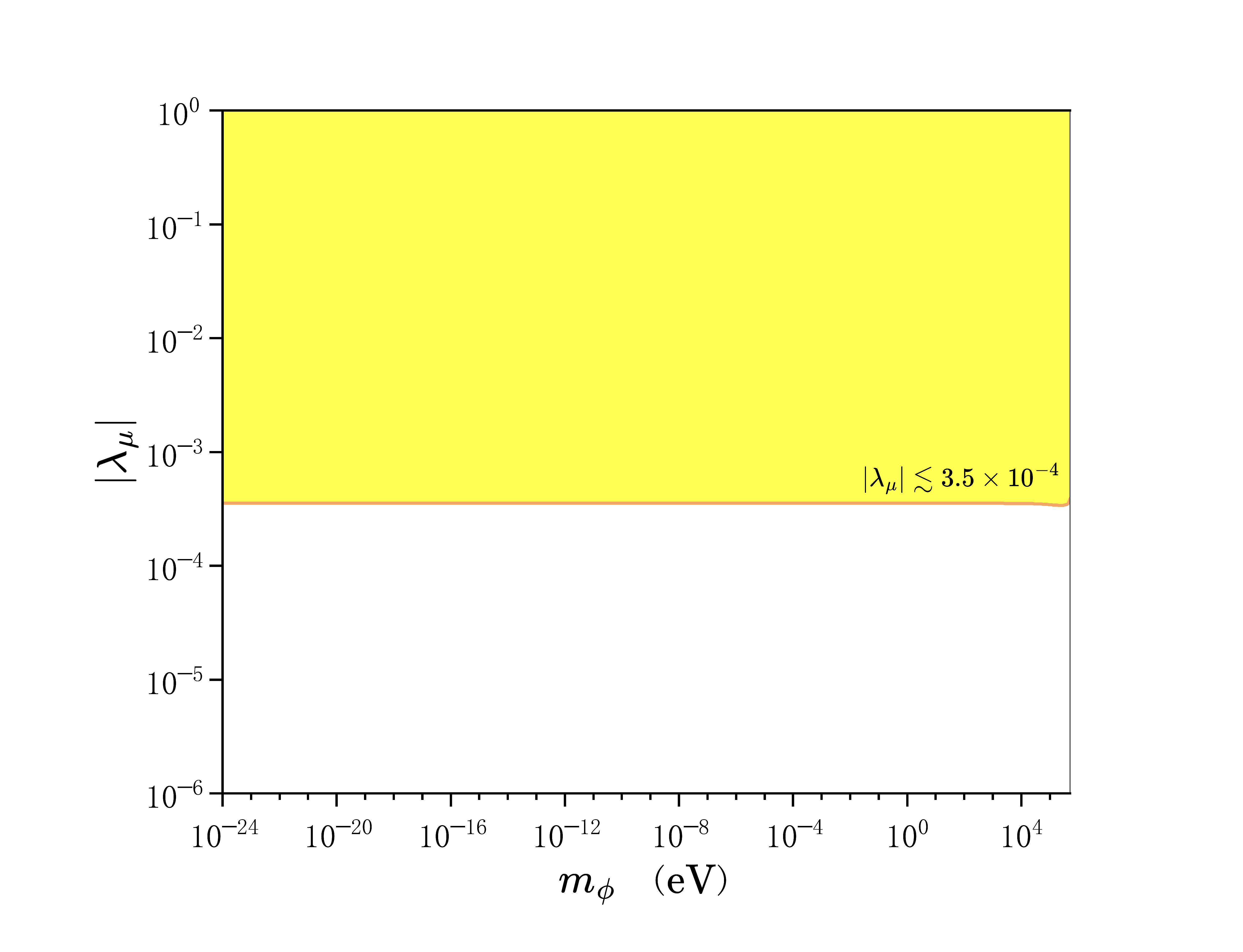}
		\includegraphics[width=0.45\textwidth]{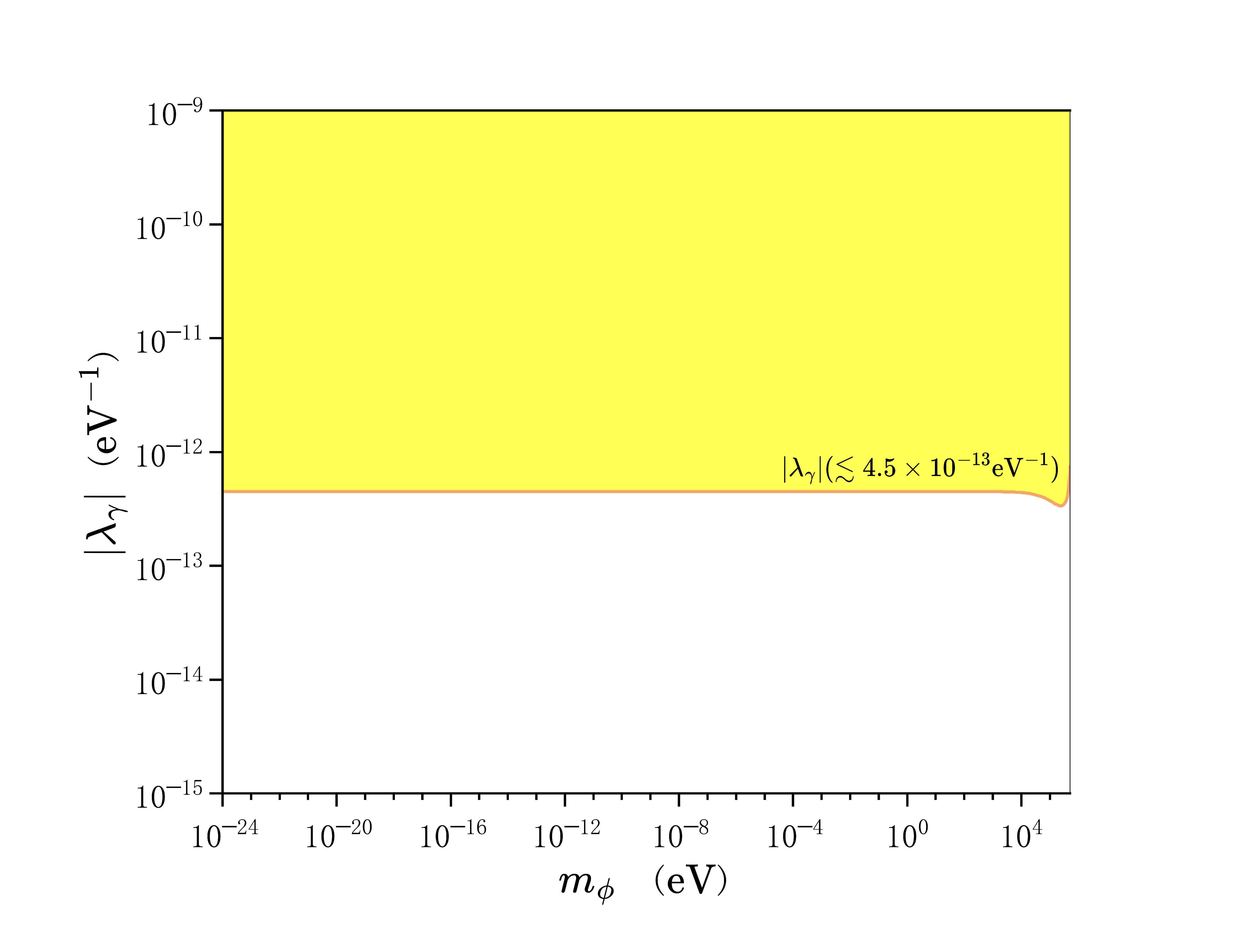}
	\caption{Full constraints on $\lambda_e$, $\lambda_\mu$, and $\lambda_\gamma$. The excluded regions are shown in yellow. For $\lambda_e$, the excluded region by the stellar cooling observations is shown in red shadow.}
	\label{fig101112}
\end{figure}

All three constraints are shown in Fig. \ref{fig101112}. One can see that, for the scalar mass below $10^4$ eV, the allowed regions for all three coupling parameters are flat straps. In other words, constraints on three coupling parameters are almost independent of the scalar mass for $m_{\phi}<10^4$ eV. For $\lambda_e$, the bound is $|\lambda_e| \leq 6.0\times 10^{-6}$. In Refs. \cite{Hardy2017, PhysRevD.96.115021}, the authors studied the constraint on $\lambda_e$ from the so-called stellar cooling observations, which turns out to be $|\lambda_e| \leq 7.0\times 10^{-16}$. Our constraint is consistent with, although not as strict as, the stellar-cooling constraint. For $\lambda_\mu$, the bound is $|\lambda_\mu| \leq 3.5\times 10^{-4}$. For $\lambda_\gamma$, the bound is $|\lambda_\gamma| \leq 4.5\times 10^{-13}$ ${\rm eV^{-1}}$. 

\subsection{Constraints on parameter pairs} 

After constraints on individual parameters are found, we continue to investigate correlations among three coupling parameters.

\begin{figure}[htbp]
	\includegraphics[width=0.5\textwidth]{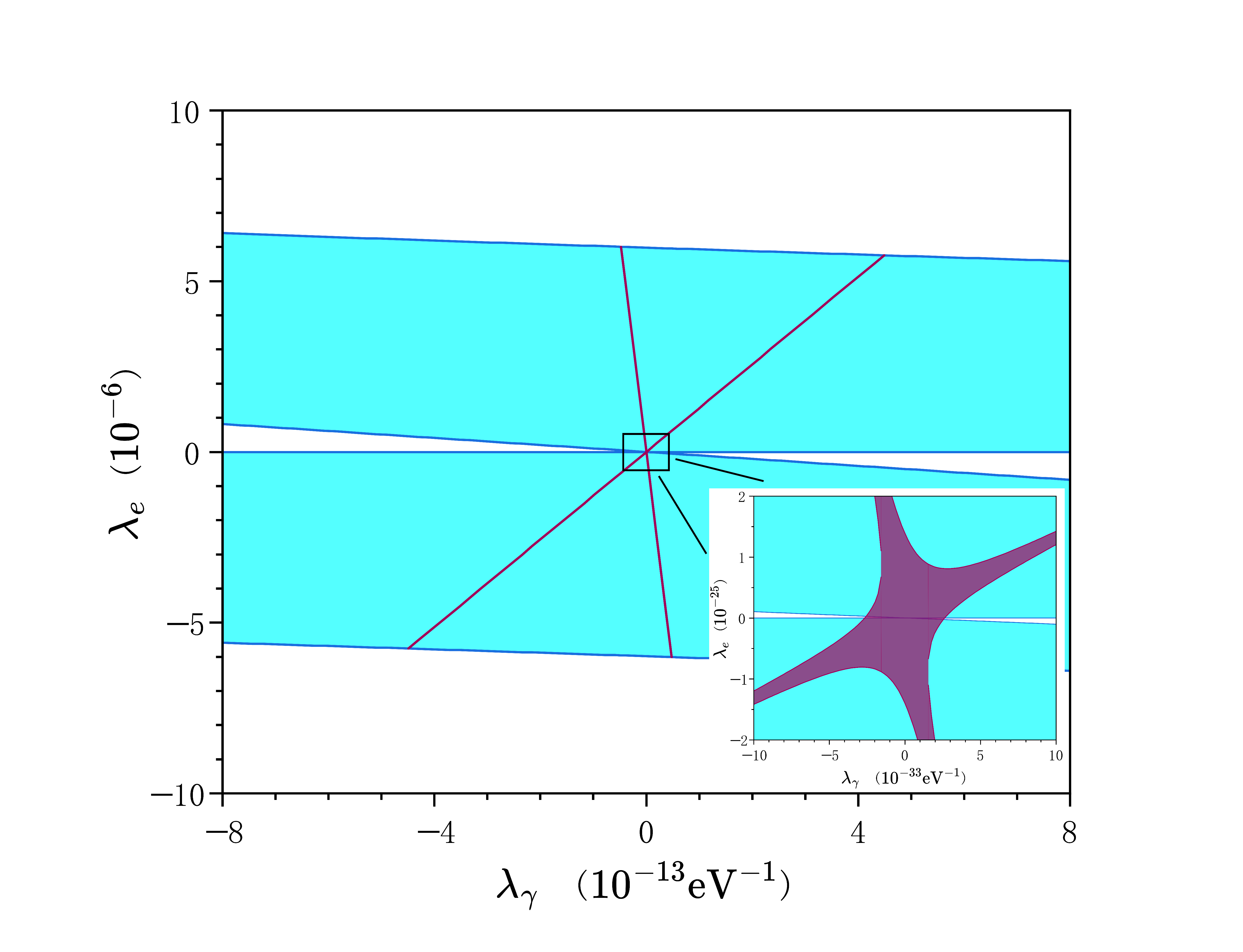}
	\includegraphics[width=0.5\textwidth]{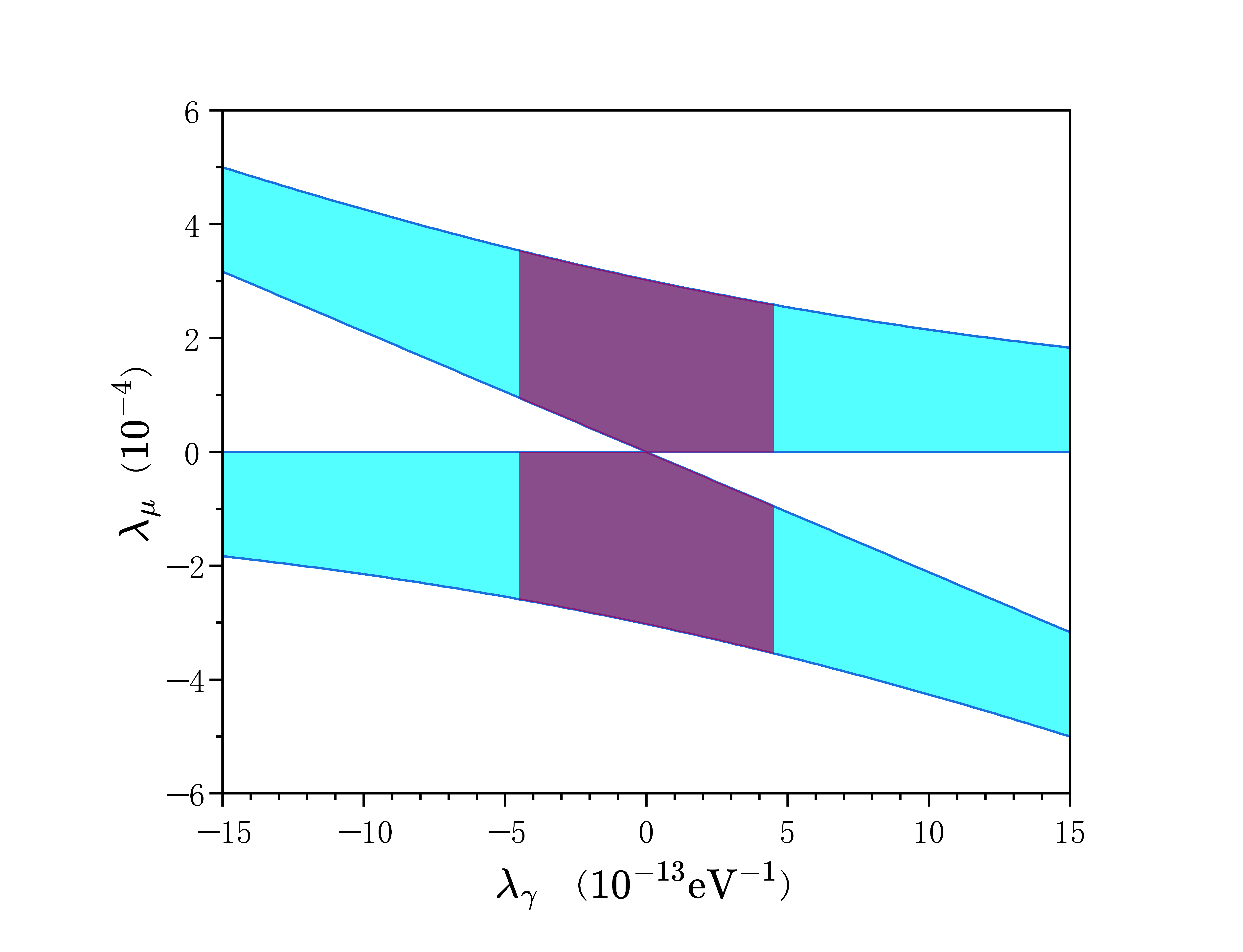}
	\caption{Constraints on the $\lambda_e$-$\lambda_\gamma$ pair, and the $\lambda_\mu$-$\lambda_\gamma$ pair. The allowed regions determined by all three experimental results are shown in violet. As comparison, the allowed regions determined by lepton anomalous magnetic moment measurements alone are shown in blue. Without losing generality, we have taken a typical value $m_\phi=10^{-14}$ eV to draw the figures.}
	\label{fig1314}
\end{figure} 
Constraints on the $\lambda_e$-$\lambda_\gamma$ and $\lambda_\mu$-$\lambda_\gamma$ pairs are shown in Fig. \ref{fig1314}. Actually, according to Eq. (\ref{deltaa}), the lepton anomalous magnetic moment measurements alone can be used to set constraints on the $\lambda_e$-$\lambda_\gamma$ and $\lambda_\mu$-$\lambda_\gamma$ pairs, which are shown as blue regions in Fig. \ref{fig1314}. Compared with blue regions which spread a large area in parameter space, the violet regions only cover finite areas. This greatly confine the allowed parameter region, which clearly shows the advantage of putting all three experimental results together to constrain the scalar model.

\begin{figure}[htbp]
\includegraphics[width=0.5\textwidth]{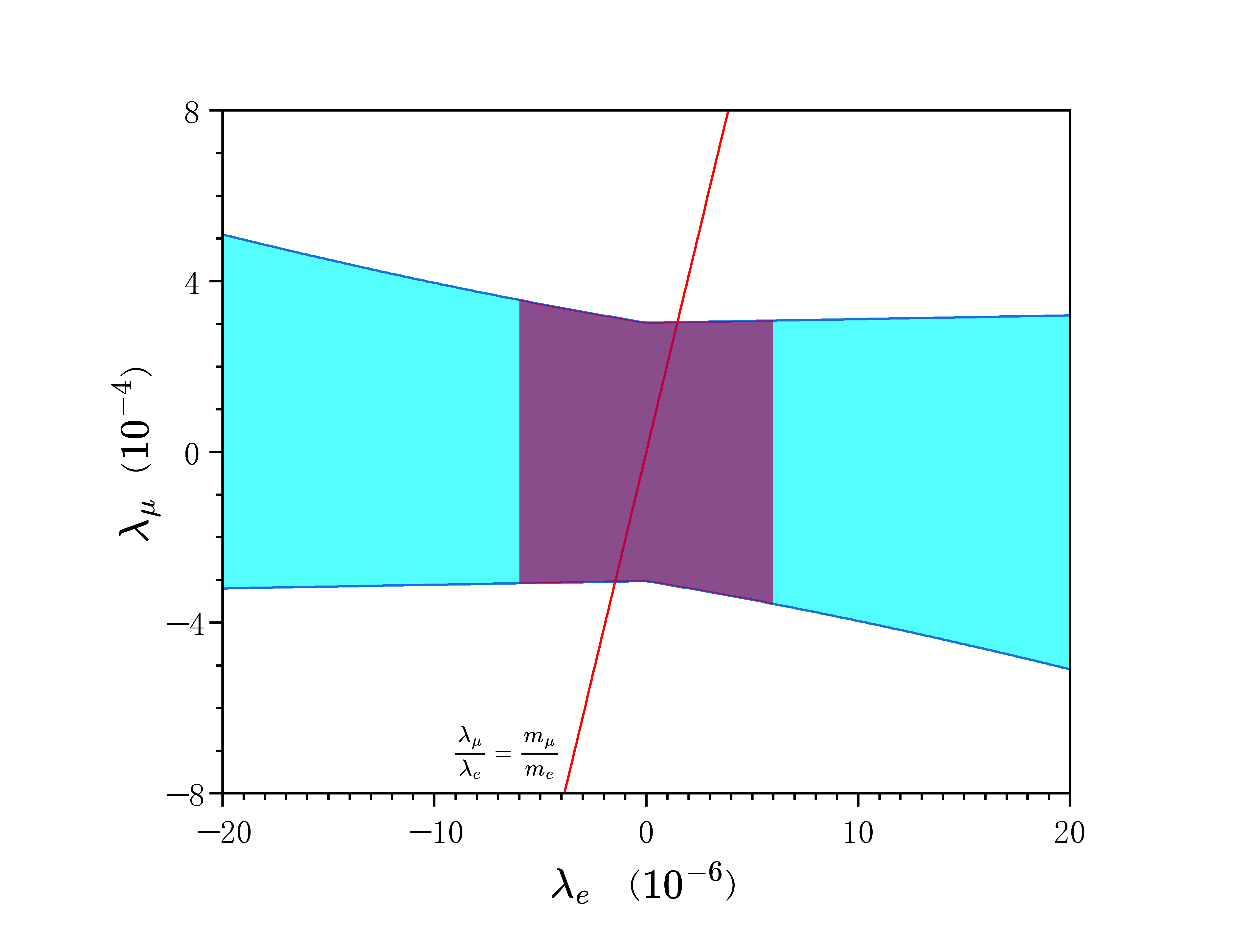}
\caption{\label{fig15} Constraint on the $\lambda_\mu$-$\lambda_e$ pair, where the violet region is the allowed region. The red line is the naive scaling value, $m_\mu/m_e$. As comparison, the allowed regions determined by lepton anomalous magnetic moment measurements alone are shown in blue. Without losing generality, we have taken a typical value $m_\phi=10^{-14}$ eV.}
\end{figure}

Another interesting issue is about the so-called naive scaling \cite{Giudice2012}. It states that contributions from new physics to lepton anomalous magnetic moments scale with the square of lepton masses. In other words, naive scaling indicates that $\delta a_\mu/\delta a_e=(m_\mu/m_e)^2$. In Eq. (\ref{deltaa}), one has $a_{see}=a_{s\mu\mu}$ and $b_{se\gamma}=b_{s\mu\gamma}$ for $m_\phi <0.1 m_e$. Then, naive scaling implies that $\lambda_\mu/\lambda_e=m_\mu/m_e$. In Fig. \ref{fig15}, the $\lambda_\mu$-$\lambda_e$ diagram is drawn. Clearly, it shows that naive scaling is favored by three experimental results.

\section{The minimal SM extension by one scalar} \label{minimalSME}

In this section, let us focus particularly on a class of scalar models, which introduce new scalars by extending the scalar sector of the SM. In the minimal extension of the SM scalar sector \cite{Patt2006fw,PhysRevD.75.037701,PhysRevD.77.035005,PhysRevD.82.043533}, it contains an additional real scalar field with no gauge quantum numbers. Such a scalar field does not couple to the SM particles directly but rather through its mixing with the Higgs field. The relevant part of the Lagrangian is the following \cite{PhysRevD.82.043533}
\begin{eqnarray*}
	V_{\varphi}= -\frac{m_h^2}{2}	H^\dagger H +\lambda_h (H^\dagger H)^2+ \mathcal{A} \varphi H^\dagger H + \frac{m_\varphi^2}{2}\varphi^2\, ,
\end{eqnarray*}	
where $m_h$ is the mass of the Higgs field, which is measured to be 125.2 GeV \cite{PhysRevD.110.030001}. $H$ stands for the Higgs field, and $\mathcal{A}$ stands for the coupling parameter between $H$ and the new scalar field $\varphi$.

 After spontaneous symmetry breaking, one has two vacuum expectation values, $\left<H^\dagger H\right>=v^2/2$ and $\left<\varphi\right>=\varphi_0$, with $v=$246 GeV. Following the notations in Ref. \cite{PhysRevD.82.043533}, the above interaction will induce some effective couplings between the scalar and the SM fields, which is 
\begin{eqnarray}\label{minimalsme}
	\mathcal{L}_{eff}&=&\frac{\mathcal{A} v}{m_h^2} \left[ g_{hff}\bar{\psi}_f\psi_f + \frac{g_{h\gamma\gamma}}{v}F_{\mu\nu}F^{\mu\nu}\right.\nonumber\\
	&&+\left. \frac{g_{hgg}}{v}F^A_{\mu\nu}F^{A\mu\nu} \right] \phi\, ,
\end{eqnarray}	
where $\phi\equiv \varphi-\varphi_0$. $g_{hff}$ stands for the Yukawa couplings of the Higgs field to the SM fermions. $g_{h\gamma\gamma}$ stands for the effective coupling of the Higgs field to the electromagnetic field. $g_{hgg}$ stands for the effective coupling of the Higgs field to the gluon field. In the SM, 
\begin{eqnarray*}
g_{hff}=m_f/v\, , \,\, g_{h\gamma\gamma}\simeq \alpha/(8\pi)\, , \,\, g_{hgg}= \alpha_s/(4\pi)\, . 
\end{eqnarray*}
Here, $\alpha_s$ is the strong coupling constant, whose numerical value is taken to be 0.5 at the running energy scale 1 GeV, as explained in Ref. \cite{PhysRevD.82.084033}.

Compared to our notations, it is easy to write down the translation between them
\begin{eqnarray}\label{minimalsmeA}
	\lambda_\gamma= \frac{\mathcal{A}}{m_h^2}\, g_{h\gamma\gamma}, \, \lambda_g= \frac{\mathcal{A}}{m_h^2}\, g_{hgg}, \, \lambda_f= \frac{\mathcal{A} v}{m_h^2}\, g_{hff}\,\,\,\,\,\,\,
\end{eqnarray}	
Since there is only one coupling parameter $\mathcal{A}$ in the model (\ref{minimalsme}),   $\lambda_\gamma$, $\lambda_e$ and $\lambda_\mu$ are not independent to each other. They should satisfy the following relations
\begin{align*}
	&\lambda_e/\lambda_\gamma = v g_{hee}/g_{h\gamma\gamma}=8\pi m_e/\alpha\, ,\\
	&\lambda_\mu/\lambda_\gamma = v g_{h\mu\mu}/g_{h\gamma\gamma}=8\pi m_\mu/\alpha\,.
\end{align*} 
One can easily see that the naive scaling, $\lambda_\mu/\lambda_e=m_\mu/m_e$, is satisfied for the model (\ref{minimalsme}). 

By inserting Eq. (\ref{minimalsmeA}) into Eqs. (\ref{deltaa}) and (\ref{etaequation1}), we get the following equations
\begin{eqnarray}
&&\delta a_{l}=\frac{\mathcal{A}^2 {m_l}^2}{m_h^4}\left( a_{sll}(r_l)+\frac{\alpha}{8\pi} b_{sl\gamma}(r_l)\right)\, ,\label{deltaa5} \\	 
&&\eta(\mathrm{Pt,Ti})=\frac{\mathcal{A}^2}{m_h^4}\left(1+\frac{R_E}{\Lambda_{\phi}}\right)I\left(\frac{R_E}{\Lambda_{\phi}}\right) e^{-R_E/\Lambda_{\phi}}\nonumber\\
	&&\times(-1.8 m_e +9.2\times10^{5}\alpha\cdot{\rm eV}\nonumber\\
	&&-1.2\times10^{8}\alpha_s\cdot{\rm eV} -39.7 m_d-39.7 m_u )\nonumber\\
	&&\times(0.21 m_e +1.1\times10^{6}\alpha\cdot{\rm eV}+3.9\times10^{5}\alpha_s\cdot{\rm eV}\nonumber\\
	&&-1.6 m_d -1.1 m_u)\times 10^{36}\, .\,\,\,\,
	\label{etaMICROSCOPE3}
\end{eqnarray}
Note that Eq. (\ref{etaMICROSCOPE3}) also contains contributions from the effective couplings of the scalar to quarks and gluons. 

\begin{figure}[htbp]
	\includegraphics[width=0.5\textwidth]{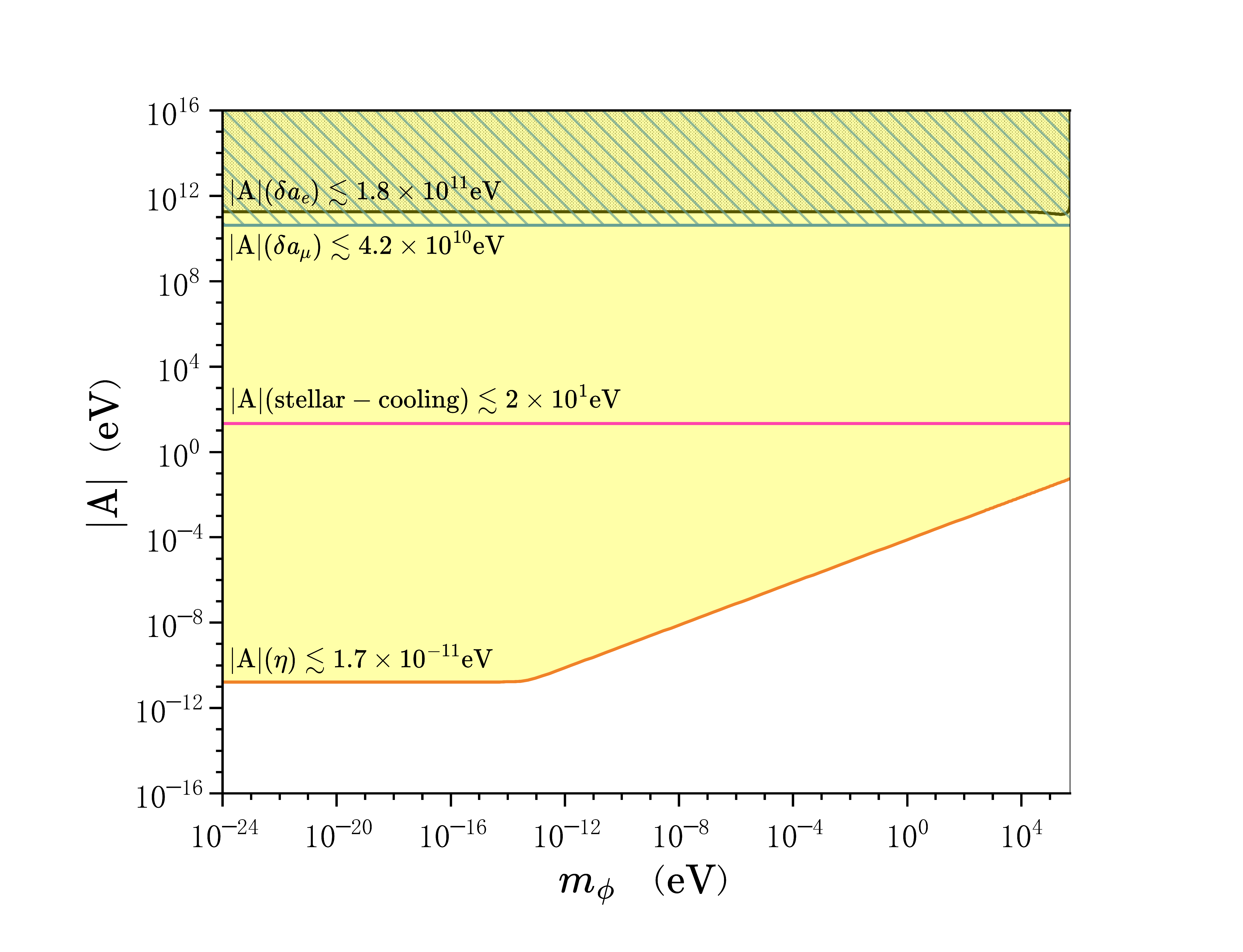}
	\caption{\label{fig18} Constraints on $\mathcal{A}$, where the regions excluded by three experimental results are drawn in various shadowed areas. For comparison, the constraint from the stellar cooling observations \cite{Hardy2017, PhysRevD.96.115021} is also shown.}
\end{figure}

Actually, with Eqs. (\ref{deltaa5}) and (\ref{etaMICROSCOPE3}), three experimental results (\ref{daelectron}), (\ref{damuon}) and (\ref{etaresult}) can set three independent bounds for $\mathcal{A}$, as shown in Fig. \ref{fig18}. The regions excluded by three experimental results are drawn in various shadowed areas. The bound from the result (\ref{etaresult}) is the best one, which is $|\mathcal{A}|\leq 1.7 \times 10^{-11}$ eV for $m_{\phi}< 10^{-13}$ eV. It is clear that the minimal SM extension by one scalar is in favor for $m_{\phi}< 10^4$ eV. For comparison, the constraint from the stellar cooling observations \cite{Hardy2017, PhysRevD.96.115021} is also shown in Fig. \ref{fig18}. Again, our constraint is consistent with the stellar-cooling constraint.

\section{Conclusion and Discussion}\label{sec:5}

The lepton anomalous magnetic moment together with the WEP violation involve all the four fundamental interactions in nature. Suppose that discrepancies between the SM predictions and measurements for the lepton (electron and muon) anomalous magnetic moments, and the WEP violation are all caused by a new scalar. By combining these three experiments together, we get new constraints on the new scalar, which could not be obtained by using either the lepton anomalous magnetic moment or the WEP violation alone. We show that the naive scaling relationship between the scalar-muon coupling and the scalar-electron coupling is favored by three experimental results. Furthermore, the model parameter of the minimal SM extension by one scalar is also constrained for $m_{\phi}<10^4$ eV.

For the muon anomalous magnetic moment, the uncertainty in $\delta a_{\mu}^{\rm EXP}(2025)$ is mainly due to the precision of the SM prediction $a_{\mu}^{\mathrm{SM}}(2025)$. To match the precision of the latest experimental average, the precision of the SM prediction has to be improved by a factor of four, which will be the main task for the next few years \cite{wp25}. Then, the role of new physics beyond the SM will be clear. For the electron anomalous magnetic moment, new measurement is ongoing to realize new precision \cite{Fan2023}. Moreover, some space-based proposals, such as STE-QUEST \cite{PhysRevD.109.064010}, plan to push the WEP test to the $10^{-17}$-level with atom interferometry. Considering all these potential progresses, we could expect to set better bounds on the new scalar in the future.

Finally, let us conmment on the tensions between experiments \cite{Parker2018} and \cite{Morel2020} in the determination of the fine-structure constant. As mentioned in the end of Sec. \ref{sec:2.1}, the new scalar $\phi$ contributes positively to $\delta a_e$ for $m_{\phi}<m_e$, and negatively to $\delta a_e$ for $m_{\phi}>2 m_e$. In this work, since we meant to focus on  new light scalars, we adopt the experiment \cite{Morel2020} to yield the positive experimental result (\ref{daelectron}). If the experiment \cite{Parker2018} is adopted, we will end up with a negative experimental result, $\delta a_e^{\rm EXP}=-1.02(26)\times{10}^{-12}$. Combining this result with results (\ref{damuon}) and (\ref{etaresult}) would shift our focus to new scalars with mass larger than 1 MeV, which will be the topic of our future work. 

\begin{acknowledgments}
This work was supported by the Technological Innovation 2030 "Quantum Communication and Quantum Computer" Major Project (Grants No. 2021ZD0300603 and No. 2021ZD0300604), the Hubei Provincial Science and Technology Major Project (Grant No. ZDZX2022000001), and the Shandong Provincial Natural Science Foundation (Grant No. ZR2023QA143).
\end{acknowledgments}



\vskip 1.2 cm

\appendix

\section{Calculation of one loop contribution to $\delta a_l$} \label{caloneloop}

\begin{figure}[htbp]
	\centering
	\includegraphics[width=0.17\textwidth]{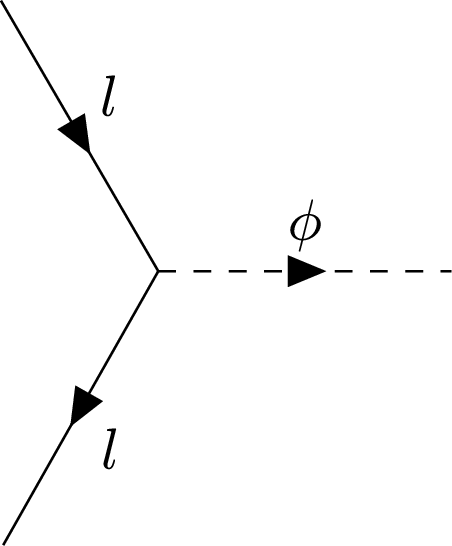}
	\includegraphics[width=0.17\textwidth]{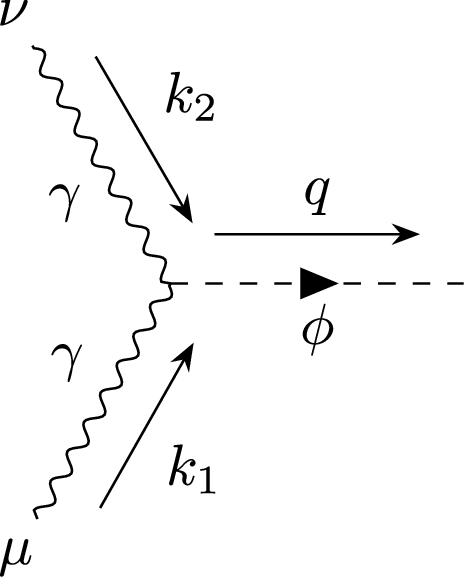}
	\caption{The left is the Scalar-Lepton-Lepton Vertex: $i\lambda_e$. The right is the Scalar-Photon-Photon Vertex: $4i \lambda_{\gamma} [{k_1}^{\nu}{k_2}^{\mu}-g^{\mu\nu}{k_1}\cdot{k_2}]$.}
	\label{fig0102}
\end{figure}
Let us start with the general linear coupling model, where linear couplings between the new scalar $\phi$ and the SM particles are assumed. Following Ref. \cite{PhysRevD.82.084033}, the general interaction terms can be written as follows,
\begin{eqnarray} \label{scalarmodel1}
	\mathcal{L}_{int}=&&\phi\left[\lambda_{\gamma} F_{\mu\nu}F^{\mu\nu}+\frac{\lambda_g \beta_3}{2g_3}{F^A}_{\mu\nu}{F^A}^{\mu\nu}\right. \nonumber\\
	&&+\left. \sum_{i=l,q}(\lambda_i+\gamma_{m_i} \lambda_g m_i)\bar{\psi}_i\psi_i\right]\, ,
\end{eqnarray}
where $\psi_l$ stands for the lepton fields for $l=e, \mu$ and $\psi_q$ stands for the quark fields for $q=u, d$. $g_{3}$ is the QCD gauge coupling, and $\beta_{3}$ is the $\beta$-function for $g_{3}$. $m_{i}$ denotes the fermionic masses (leptons and quarks). $\gamma_{m_{i}}$ is the anomalous dimension due to the renormalization-group running of the quark masses.  $\lambda_{\gamma}$ and $\lambda_g$ denote the couplings to the $U(1)$ photon and the $SU(3)$ gluons, respectively. $\lambda_l$ denotes the dimensionless Yukawa coupling to leptons, and $\lambda_q$ denotes the dimensionless Yukawa coupling to quarks.  
In total, there are six coupling parameters ($\lambda_\mu$,$\lambda_e$,$\lambda_\gamma$,$\lambda_u$,$\lambda_d$, and $\lambda_g$).
Note that the relation between notations used in Ref. \cite{PhysRevD.82.084033} and ours is: $\lambda_{\gamma} \equiv \kappa d_e$, $\lambda_g \equiv -\kappa d_g$, $\lambda_i \equiv -\kappa m_i d_{m_i}$, where $\kappa \equiv \sqrt{4\pi G}$. 
The corresponding Feynman rules are given in Fig. \ref{fig0102}.

As discussed in \cite{Nowakowski2005}, one can obtain various form factors $F_i$'s extracted from the lepton-photon vertex $\Gamma^{\mu}=\gamma^{\mu}F_1(q^2)+\frac{i\sigma^{\mu\nu}}{2m_l}q_{\nu}F_2(q^2)+\frac{i\sigma^{\mu\nu}}{2m_l}q_{\nu}\gamma_5 F_3(q^2)+\frac{1}{2m_l}(q^{\mu}-\frac{q^2}{2m_l}\gamma^{\mu})\gamma_5 F_4(q^2)$, where $q$ is the momentum of the external photon with on-shell condition $q^2=0$. The lepton anomalous magnetic moment $a_l$ is defined to be
\begin{eqnarray}
	a_l &\equiv& {\rm{Re}}(F_2(0)),\nonumber\\
	\delta a_l&\equiv& a_l-a_l^{{\rm SM}} ,
\end{eqnarray}
where the SM prediction $a_l^{{\rm SM}}$ has been calculated in many studies (for example, see Refs. \cite{PhysRevLett.109.111807,atoms7010028,volkov2024} for the electron, and Refs. \cite{PhysRevLett.109.111808,AOYAMA20201,colangelo2022prospects} for the muon).

\begin{figure}[htbp]
	\centering
	\includegraphics[width=0.2\textwidth]{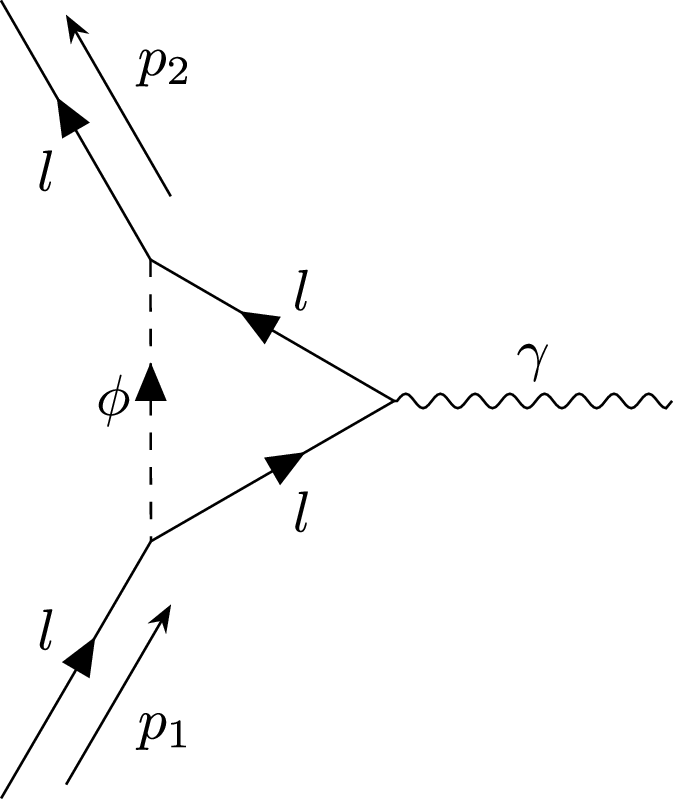}
	\caption{\label{fig3} The Scalar-Lepton-Lepton loop diagram}
\end{figure}

\begin{figure}[htbp]
	\centering
	\includegraphics[width=0.36\textwidth]{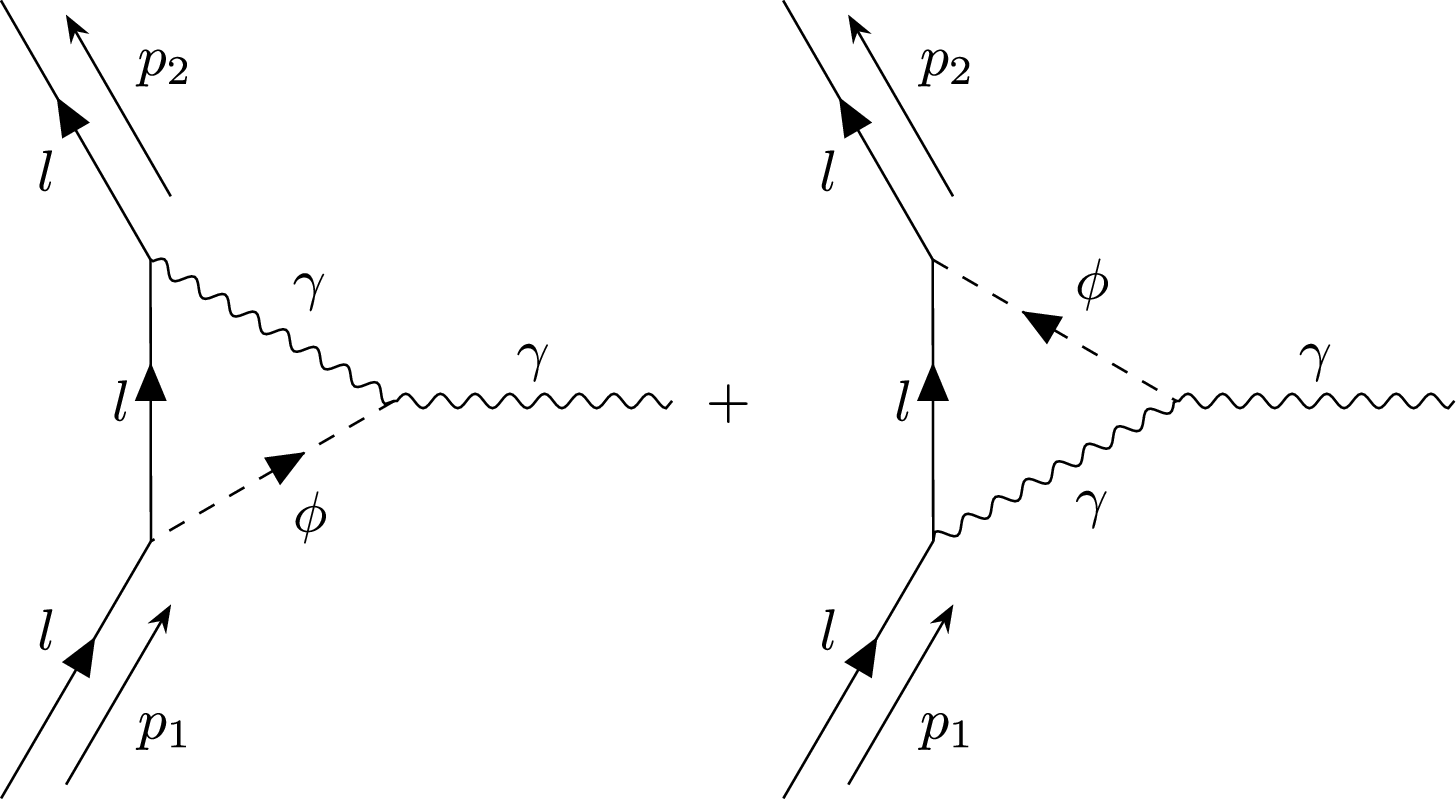}
	\caption{\label{fig5} The Scalar-Lepton-Photon loop diagrams}
\end{figure}

At one-loop level, two types of Feynman diagrams contribute to $\delta a_l$: the Scalar-Lepton-Lepton loop diagram (Fig. \ref{fig3}), and the Scalar-Lepton-Photon loop diagram (Fig. \ref{fig5}). We denote contributions of the former diagram as $\delta a_{l1}$, and the latter as $\delta a_{l2}$. Then, we have
\begin{equation}
	\delta a_l = \delta a_{l1}+\delta a_{l2}\equiv {\rm{Re}}(F_2^{(1)}(0)+F_2^{(2)}(0)),
\end{equation}
where $F_2^{(i)}$ stands for the form factors of the two diagrams.

Let us first calculate the Scalar-Lepton-Lepton loop diagram (Fig. \ref{fig3}), using the Passarino-Veltman Renormalization \cite{PASSARINO1979151,THOOFT1979365}. With the help of FeynCalc \cite{MERTIG1991345, SHTABOVENKO2020107478} and FeynArts \cite{HAHN2001418}, we can derive the amplitude of Fig. \ref{fig3} :
\begin{widetext}
	\begin{align}\label{amp01}
		&{\rm Amp01}= 2 i \pi ^2 {e}{\lambda_l}^2 m_l (\text{p}_1^{\mu }+\text{p}_2^{\mu }) \left[2 \text{C}_1(m_l^2,0,m_l^2,m_{\phi}^2,m_l^2,m_l^2)+\text{C}_{11}(m_l^2,0,m_l^2,m_{\phi}^2,m_l^2,m_l^2)\right.\\  \nonumber
		&\left.+\text{C}_{12}(m_l^2,0,m_l^2,m_{\phi}^2,m_l^2,m_l^2)\right] (\varphi(\text{p}_2,m_l))\cdot(\varphi(\text{p}_1,m_l)) -i \pi^2 {e}{\lambda_l}^2 \left[\text{B}_0(0,m_l^2,m_l^2)+\right.\\  \nonumber
		&\left. (m_{\phi}^2-4 m_l^2) \text{C}_0(0,m_l^2,m_l^2,m_l^2,m_l^2,m_{\phi}^2) -2 \text{C}_{00}(m_l^2,0,m_l^2,m_{\phi}^2,m_l^2,m_l^2)\right] (\varphi(\text{p}_2,m_l))\cdot\gamma ^{\mu }\cdot(\varphi(\text{p}_1,m_l)) ,
	\end{align}
\end{widetext}
where $\varphi(\text{p}_i,m_e)$ stands for the electron field,  $p_1$ is the incoming momentum of the lepton and $p_2$ is the outcoming momentum. The involved one-point, two-point and three-point Passarino-Veltman coefficient functions are defined as follows
\begin{widetext}
	\begin{align}
		&\text{A}_{0}(m_0^2)= {\bm{\mu}}^{2\epsilon} \int \frac{d^d k}{(2\pi)^d} \frac{1}{k^2-m_0^2+i\varepsilon}\, ,\\
		&\text{B}_{\{0,\mu,\mu\nu \}}(p^2;m_0^2,m_i^2)= {\bm{\mu}}^{2\epsilon} \int \frac{d^d k}{(2\pi)^d}  \frac{\{1, k_{\mu},  k_{\mu}k_{\nu}\}}{(k^2-m_0^2+i\varepsilon)((k+p)^2-m_i^2+i\varepsilon)}\, ,\\
		&\text{C}_{\{0,\mu,\mu\nu \}}(p_1^2,q^2,p_2^2;m_0^2,m_1^2,m_2^2)= \int \frac{d^d k}{(2\pi)^d}  \frac{ \{1, k_{\mu},  k_{\mu}k_{\nu}\}{\bm{\mu}}^{2\epsilon}}{(k^2-m_0^2+i\varepsilon)((k+p_1)^2-m_1^2+i\varepsilon)((k+p_2)^2-m_2^2+i\varepsilon)} \, ,\\
		&\text{C}_{\mu}={p_1}_{\mu}\text{C}_1+{p_2}_{\mu}\text{C}_2\, ,\\
		&\text{C}_{\mu\nu}=g_{\mu\nu}\text{C}_{00}+{p_1}_{\mu}{p_1}_{\nu}\text{C}_{11}+{p_2}_{\mu}{p_2}_{\nu}\text{C}_{22}+({p_1}_{\mu}{p_2}_{\nu}+{p_2}_{\mu}{p_1}_{\nu})\text{C}_{12}\, ,
	\end{align}
\end{widetext}
where $\{0,\mu,\mu\nu \}$ stands for an index being $0$, $\mu$, or $\mu\nu$ which corresponds to the momentum $\{1, k_{\mu},  k_{\mu}k_{\nu}\}$ being $1$, $k_{\mu}$, or $k_{\mu}k_{\nu}$. $\varepsilon$ is the infinitesimal in Feynman prescription of pole. $\bm{\mu}$ is the 't Hooft parameter as a mass parameter introduced through dimensional regularization. $\epsilon\equiv(4-d)/2$ is the dimension regulated under the dimensional regularization. 

$F_2^{(1)}(0)$ can be extracted from the piece proportional to $(p_1^{\mu }+p_2^{\mu })$ in Eq. (\ref{amp01}), which yields
\begin{align}
	&F_2^{(1)}(0)=4 \pi ^2 {\lambda_l}^2 m_l^2 \left[2 \text{C}_1(m_l^2,0,m_l^2,m_{\phi}^2,m_l^2,m_l^2)\right.\nonumber\\ &\left.+\text{C}_{11}(m_l^2,0,m_l^2,m_{\phi}^2,m_l^2,m_l^2)\right.\nonumber\\	&\left.+\text{C}_{12}(m_l^2,0,m_l^2,m_{\phi}^2,m_l^2,m_l^2)\right].\nonumber
\end{align}
Coefficient functions ($C_1$, $C_{11}$ and $C_{12}$) can be evaluated with $Mathematica$ packages, such as PackageX \cite{PATEL2015276, PATEL201766}. Thus, we get
\begin{widetext}
	\begin{align}
	&F_2^{(1)}(0)=-\frac{1}{16\pi^2m_l^4}{\lambda_l}^2\left[m_{\phi}^4\log(\frac{m_{\phi}^2}{m_l^2})-2 m_{\phi}^2\sqrt{m_{\phi}^4-4m_{\phi}^2m_l^2}
		\log(\frac{m_{\phi}^2+\sqrt{m_{\phi}^4-4m_{\phi}^2m_l^2}}{2m_{\phi} m_l})+3m_l^4 \right.\nonumber\\
		&\left.+m_l^2\left(-3m_{\phi}^2\log(\frac{m_{\phi}^2}{m_l^2}) -2m_{\phi}^2+2\sqrt{m_{\phi}^4-4m_{\phi}^2m_l^2}\log(\frac{m_{\phi}^2+\sqrt{m_{\phi}^4-4m_{\phi}^2m_l^2}}{2m_{\phi} m_l})\right)\right]\, ,
	\end{align}
\end{widetext}
which is consistent with the result in Ref. \cite{PhysRevD.93.035006}. Then, we find its contribution to $\delta a_l$, which is
\begin{equation}
	\delta a_{l1}(m_{\phi})={\rm{Re}}(F_2^{(1)}(0))\equiv{\lambda_l}^2a_{sll}(r_l)\, ,
\end{equation}
with
\begin{widetext}
\begin{eqnarray}
		a_{sll}(r_l)=
		\begin{cases}
			\frac{-2{r_l}^2-3{r_l}^2\log({r_l}^2)+{r_l}^4\log({r_l}^2)-2\sqrt{4{r_l}^2-{r_l}^4}{r_l}^2\cos^{-1}(\frac{r_l}{2})+2\sqrt{4{r_l}^2-{r_l}^4}\cos^{-1}(\frac{r_l}{2})+3}{16\pi^2} & \text{if } r_l \leq 2,\,\,\,\,\,\,\,\\
			
			\frac{-2{r_l}^2-3{r_l}^2\log({r_l}^2)+{r_l}^4\log({r_l}^2)-2\sqrt{{r_l}^4-4{r_l}^2}{r_l}^2\cosh^{-1}(\frac{r_l}{2})+2\sqrt{{r_l}^4-4{r_l}^2}\cosh^{-1}(\frac{r_l}{2})+3}{16\pi^2} & \text{if } r_l \geq 2 ,
		\end{cases}
\end{eqnarray}
\end{widetext}
where $r_l\equiv m_{\phi}/m_l$  with $l=e,\mu$. 

Next, let us calculate the contribution from the Sacalar-Lepton-Photon loop diagram (Fig. \ref{fig5}).
Similar to the first diagram, its amplitude is calculated to be
\begin{widetext}
	\begin{align}\label{amp02}
	&	{\rm Amp02}=i \pi ^2 {e} {\lambda_l} {\lambda_{\gamma}}(\text{p}_1^{\mu }+\text{p}_2^{\mu })  \left[ \text{B}_0(m_l^2,0,m_l^2)+m_{\phi}^2 \text{C}_0(0,m_l^2,m_l^2,m_{\phi}^2,0,m_l^2)-2 m_l^2 \text{C}_0(0,m_l^2,m_l^2,m_{\phi}^2,0,m_l^2) \right.\nonumber\\  
		& -2 m_l^2 \text{C}_1(m_l^2,0,m_l^2,m_l^2,m_{\phi}^2,0) -2 m_l^2 \text{C}_2(m_l^2,0,m_l^2,m_l^2,0,m_{\phi}^2) -2 m_l^2 \text{C}_{12}(m_l^2,0,m_l^2,m_l^2,0,m_{\phi}^2) \nonumber\\  
		&\left. - 2 m_l^2 \text{C}_{12}(m_l^2,0,m_l^2,m_l^2,m_{\phi}^2,0) \right]  (\varphi(\text{p}_2 , m_l ))\cdot(\varphi(\text{p}_1 , m_l ))   \nonumber\\
		&+i \pi ^2 {e} {\lambda_l} {\lambda_{\gamma}}\left[m_{\phi}^2( \text{p}_1^{\mu }\text{C}_1(m_l^2,0,m_l^2,m_l^2,m_{\phi}^2,0)+ \text{p}_2^{\mu } \text{C}_2(m_l^2,0,m_l^2,m_l^2,0,m_{\phi}^2)) -4 m_l^2 \text{p}_1^{\mu } \text{C}_1(m_l^2,0,m_l^2,m_l^2,0,m_{\phi}^2)  \right.   \nonumber\\
		& -4 m_l^2 \text{p}_2^{\mu } \text{C}_2(m_l^2,0,m_l^2,m_l^2,m_{\phi}^2,0) -2 m_l^2 \text{p}_1^{\mu } \text{C}_{11}(m_l^2,0,m_l^2,m_l^2,0,m_{\phi}^2)-2 m_l^2  \text{p}_1^{\mu } \text{C}_{11}(m_l^2,0,m_l^2,m_l^2,m_{\phi}^2,0)   \nonumber\\
		&\left. - 2 m_l^2 \text{p}_2^{\mu } \text{C}_{22}(m_l^2,0,m_l^2,m_l^2,0,m_{\phi}^2) -2 m_l^2 \text{p}_2^{\mu } \text{C}_{22}(m_l^2,0,m_l^2,m_l^2,m_{\phi}^2,0) \right](\varphi(\text{p}_2 , m_l ))\cdot(\varphi(\text{p}_1 , m_l ))  \nonumber\\
		&-\frac{i \pi ^2 {e} {\lambda_l} {\lambda_{\gamma}} }{2 m_l}\left[4 m_l^2 \text{B}_0(m_l^2,m_{\phi}^2,m_l^2)-2 m_l^2 m_{\phi}^2 \text{C}_0(0,m_l^2,m_l^2,m_{\phi}^2,0,m_l^2)- m_l^2 m_{\phi}^2 \text{C}_2(m_l^2,0,m_l^2,m_l^2,m_{\phi}^2,0) \right.   \nonumber\\
		& - m_l^2 m_{\phi}^2 \text{C}_1(m_l^2,0,m_l^2,m_l^2,0,m_{\phi}^2)+ m_l^2 m_{\phi}^2 \text{C}_1(m_l^2,0,m_l^2,m_l^2,m_{\phi}^2,0)+ m_l^2 m_{\phi}^2 \text{C}_2(m_l^2,0,m_l^2,m_l^2,0,m_{\phi}^2)    \nonumber\\
		&\left. - m_{\phi}^2 \text{B}_0(m_l^2,m_{\phi}^2,m_l^2) +4 m_l^2( \text{C}_{00}(m_l^2,0,m_l^2,m_l^2,0,m_{\phi}^2)+ \text{C}_{00}(m_l^2,0,m_l^2,m_l^2,m_{\phi}^2,0))-2 \text{A}_0(m_l^2)+ \text{A}_0(m_{\phi}^2)\right] \nonumber\\
		&\times (\varphi  ( \text{p}_2 , m_l ))\cdot\gamma ^{\mu }\cdot(\varphi  ( \text{p}_1 , m_l )) 
	\end{align}
$F_2^{(2)}(0)$ can be extracted from the piece proportional to $(p_1^{\mu }+p_2^{\mu })$ in Eq. (\ref{amp02}), which turns out to be
	\begin{align}
		&\frac{i{e}}{2 m_l}(\text{p}_1^{\mu }+\text{p}_2^{\mu }) F_2^{(2)}(0)=i \pi ^2 {e} {\lambda_l} {\lambda_{\gamma}}(\text{p}_1^{\mu }+\text{p}_2^{\mu })  \left[ \text{B}_0(m_l^2,0,m_l^2)+m_{\phi}^2 \text{C}_0(0,m_l^2,m_l^2,m_{\phi}^2,0,m_l^2) \right.\\  \nonumber
		&-2 m_l^2 \text{C}_0(0,m_l^2,m_l^2,m_{\phi}^2,0,m_l^2) -2 m_l^2 \text{C}_1(m_l^2,0,m_l^2,m_l^2,m_{\phi}^2,0) -2 m_l^2 \text{C}_2(m_l^2,0,m_l^2,m_l^2,0,m_{\phi}^2) \\  \nonumber
		&\left. -2 m_l^2 \text{C}_{12}(m_l^2,0,m_l^2,m_l^2,0,m_{\phi}^2)- 2 m_l^2 \text{C}_{12}(m_l^2,0,m_l^2,m_l^2,m_{\phi}^2,0) \right]  (\varphi(\text{p}_2 , m_l ))\cdot(\varphi(\text{p}_1 , m_l )) \\  \nonumber
		&+i \pi ^2 {e} {\lambda_l} {\lambda_{\gamma}}\left[m_{\phi}^2 \text{p}_1^{\mu }\text{C}_1(m_l^2,0,m_l^2,m_l^2,m_{\phi}^2,0)+m_{\phi}^2 \text{p}_2^{\mu } \text{C}_2(m_l^2,0,m_l^2,m_l^2,0,m_{\phi}^2)  \right. \\  \nonumber
		& -4 m_l^2 \text{p}_2^{\mu } \text{C}_2(m_l^2,0,m_l^2,m_l^2,m_{\phi}^2,0) -2 m_l^2 \text{p}_1^{\mu } \text{C}_{11}(m_l^2,0,m_l^2,m_l^2,0,m_{\phi}^2)-2 m_l^2  \text{p}_1^{\mu } \text{C}_{11}(m_l^2,0,m_l^2,m_l^2,m_{\phi}^2,0) \\  \nonumber
		&\left. - 2 m_l^2 \text{p}_2^{\mu } \text{C}_{22}(m_l^2,0,m_l^2,m_l^2,0,m_{\phi}^2) -2 m_l^2 \text{p}_2^{\mu } \text{C}_{22}(m_l^2,0,m_l^2,m_l^2,m_{\phi}^2,0)-4 m_l^2 \text{p}_1^{\mu } \text{C}_1(m_l^2,0,m_l^2,m_l^2,0,m_{\phi}^2) \right] \\  \nonumber
	\end{align}
After coefficient functions are evaluated with PackageX \cite{PATEL2015276, PATEL201766}, we get
	\begin{align}
		&F_2^{(2)}(0)=-\frac{1}{48\pi^2 m_l^3}\lambda_l\lambda_{\gamma}\left[m_{\phi}^4\log{(\frac{m_{\phi}^2}{m_l^2})}-2m_{\phi}^2\sqrt{m_{\phi}^4-4m_{\phi}^2m_l^2}
		\log{(\frac{m_{\phi}^2+\sqrt{m_{\phi}^4-4m_{\phi}^2m_l^2}}{2m_{\phi}m_l})}\right.\\  \nonumber
		&\left. -m_l^2\left(6m_{\phi}^2\log{(\frac{m_{\phi}^2}{m_l^2})}+2m_{\phi}^2-8\sqrt{m_{\phi}^4-4m_{\phi}^2m_l^2}\log{(\frac{m_{\phi}^2+\sqrt{m_{\phi}^4-4m_{\phi}^2m_l^2}}{2m_{\phi}m_l})}\right)+6m_l^4\left(\log{(\frac{{\bm{\mu}}^2}{m_l^2})}+3+\frac{1}{\bar{\epsilon}}\right)\right]\, ,
	\end{align}
\end{widetext}
where $\frac{1}{\bar{\epsilon}}\equiv \frac{1}{\epsilon }-{\bm{\gamma}} +\log (4 \pi )$ with the Euler's constant $\bm{\gamma}$. The term, $-\frac{\lambda_l\lambda_{\gamma} m_l}{8\pi^2}\frac{1}{\bar{\epsilon}}$, is the regularized UV divergence which can be cancelled at low energy.  The IR part, $-\frac{\lambda_l\lambda_{\gamma} m_l}{8\pi^2} \log{(\frac{{\bm{\mu}}^2}{m_l^2})}$, can be cancelled by the bremsstrahlung effect. In the end, we get the second contribution to $\delta a_l$,
\begin{equation}
	\delta a_{l2}(m_{\phi})={\rm{Re}}(F_2^{(2)}(0))\equiv \lambda_l\lambda_{\gamma} m_{l}b_{sl\gamma}(r_l),
\end{equation}
where
\begin{widetext}
	\begin{eqnarray}
		b_{sl\gamma}(r_l)=
		\begin{cases}
			\frac{-2{r_l}^2-6{r_l}^2\log({r_l}^2)+{r_l}^4\log({r_l}^2)-2\sqrt{4{r_l}^2-{r_l}^4}{r_l}^2\cos^{-1}(\frac{r_l}{2})+8\sqrt{4{r_l}^2-{r_l}^4}\cos^{-1}(\frac{r_l}{2})+18}{48\pi^2} & \text{if } r_l \leq 2,\,\,\,\,\,\,\,\\
			
			\frac{-2{r_l}^2-6{r_l}^2\log({r_l}^2)+{r_l}^4\log({r_l}^2)-2\sqrt{{r_l}^4-4{r_l}^2}{r_l}^2\cosh^{-1}(\frac{r_l}{2})+8\sqrt{{r_l}^4-4{r_l}^2}\cosh^{-1}(\frac{r_l}{2})+18}{48\pi^2} & \text{if } r_l \geq 2.
		\end{cases}
	\end{eqnarray}
\end{widetext}

Then, the total one-loop contribution of the scalar field to $\delta a_{l}$ is 
\begin{equation}\label{deltaa1}
	\delta a_{l}=\delta a_{l1}+\delta a_{l2}={\lambda_l}^2 a_{sll}(r_l)+\lambda_l \lambda_{\gamma} m_{l} b_{sl\gamma}(r_l)\, .
\end{equation}
One can see that $\lambda_l$ appears in both terms, and $\lambda_{\gamma}$ appears only in the second term. Note that, at one-loop level, $\delta a_{l}$ does not get contributions from the scalar-quark coupling $\lambda_{q}$ and the scalar-gluon coupling $\lambda_{g}$.

\section{Calculation of contribution of the scalar $\phi$ to $\eta$} \label{caleta}

For the general linear coupling model (\ref{scalarmodel1}), the $\phi$ contribution to the Eötvös parameter has been calculated in Refs. \cite{PhysRevD.82.084033,Zhao2022}. Here, we quote their results as following. It is straightforward to check that the Newtonian interaction between a mass A and a mass B will be modified into the form
\begin{eqnarray}\label{vpotential}
	V=-\frac{G m_A m_B}{r_{AB}}	(1+\zeta_A\zeta_B e^{-r_{AB}/\Lambda_{\phi}}),\,\,\,\,
\end{eqnarray}
where $G$ is the Newtonian constant. $\Lambda_{\phi}\equiv\hbar/m_{\phi}$ is the Compton wavelength of the scalar $\phi$. 

$\zeta_{A}$ is the so-called scalar-charge for a mass. The scalar model leads to the $\phi$-dependence for lepton and quark masses. Ordinary matter is made of atoms, which can be further decomposed into fundamental particles (photons, electrons, gluons and quarks). Thus, the scalar model (\ref{scalarmodel}) leads to the $\phi$-dependence for atomic mass, which gives the definition for $\zeta_A$,
\begin{align} 
&\zeta_A = -\kappa^{-1} \left[\lambda_g + \frac{1}{m_A} \left( (\lambda_{\hat{m}}-\lambda_g \hat{m})\frac{\partial m_A}{\partial \hat{m}}-4\lambda_{\gamma}\alpha\frac{\partial m_A}{\partial \alpha} \right.\right.\nonumber\\
&\left.\left. + (\lambda_{\delta m}-\lambda_g \delta m)\frac{\partial m_A}{\partial \delta m} -4\lambda_{\gamma}\alpha\frac{\partial m_A}{\partial \alpha}\right.\right.\nonumber\\
&\left.\left.+ (\lambda_{m_e}-\lambda_g m_e)\frac{\partial m_A}{\partial m_e} \right) \right]\, ,
\end{align}	
where $\hat{m}=\frac{m_d+m_u}{2}, \delta m=m_d-m_u$,	$\lambda_{\hat{m}}=\frac{\lambda_d+\lambda_u}{2}$, and $\lambda_{\delta m}=\lambda_d-\lambda_u$. 

The calculation of $\zeta_{A}$ is quite complicated, which has been done in Ref. \cite{PhysRevD.82.084033},
\begin{align}\label{atomcharge}
&	\zeta_A = -\kappa^{-1} \left[ \lambda_g + (\frac{\lambda_{\hat{m}}}{\hat{m}}-\lambda_g)Q_{\hat{m}}+(\frac{\lambda_{\delta m}}{\delta m}-\lambda_g)Q_{\delta m} \right.\nonumber\\ 
	&\left.\,\,\, +(\frac{\lambda_e}{m_e}-\lambda_g)Q_{m_e} -4\lambda_{\gamma}Q_e \right] \nonumber\\ 
&=-\kappa^{-1} \left[ (1-Q_{\hat{m}}-Q_{\delta m}-Q_{m_e})\lambda_g +\frac{\lambda_e}{m_e}Q_{m_e}  \right.\nonumber\\  
	&\left. -4\lambda_{\gamma}Q_e+(\frac{Q_{\hat{m}}}{m_d+m_u}+\frac{Q_{\delta m}}{m_d-m_u})\lambda_d\right.\nonumber\\ 
	&\left.	+(\frac{Q_{\hat{m}}}{m_d+m_u}-\frac{Q_{\delta m}}{m_d-m_u})\lambda_u \right] 
\end{align}	
where
\begin{subequations}
	\begin{align}
		Q_{\hat m}=&F_A\Bigg[0.093-\frac{0.036}{A^{1/3}}-0.02\frac{(A-2Z)^2}{A^2}\nonumber\\ \label{q14a}
		&\qquad -1.4\times 10^{-4}\frac{Z(Z-1)}{A^{4/3}}\Bigg]\\ 
		Q_{\delta m}=&F_A\Bigg[0.0017\frac{A-2Z}{A}\Bigg] \\
		Q_{m_e}=&F_A\Bigg[5.5\times 10^{-4}\frac{Z}{A}\Bigg]\, ,
	\end{align}
	and
	\begin{equation}
		Q_e=F_A\Bigg[-1.4+8.2 \frac{Z}{A}+7.7 \frac{Z(Z-1)}{A^{4/3}}\Bigg]\times 10^{-4}\, . \label{q14d}
	\end{equation}
\end{subequations}
$Z$ is the atomic number, and $A$ is the mass number of atoms. The factor $F_A$ can be replaced by one in lowest approximation. 

For two test bodies freely falling towards the Earth, the Eötvös parameter $\eta$ is found to be \cite{Zhao2022}
\begin{eqnarray}\label{etaequation1}
	\eta=&&\left(1+\frac{R_E}{\Lambda_{\phi}}\right)I\left(\frac{R_E}{\Lambda_{\phi}}\right)(\zeta_{A}-\zeta_{B})\zeta_E e^{-R_E/\Lambda_{\phi}}\, , \nonumber\\
	&&	I(x)\equiv\frac{3(x\cosh{(x)}-\sinh{(x)})}{x^3}\, ,
\end{eqnarray}
where $R_E$ is the radius of the Earth. Here, the factor $I(x)$ takes into account the fact that the Earth is a sphere of finite size. 

According to Ref. \cite{pnas.77.12.6973}, the Earth is made of 49.83\% Oxygen, 15.19\% Iron, 15.14\% Magnesium, 14.23\% Silicon, 2.14\% Sulfur, 1.38\% Aluminum and 1\% Calcium. Then, one can calculate the scalar-charge of the Earth,
\begin{align}\label{earthcharge1}
	&\zeta_{\rm E}=-1.808\times10^{18}\lambda_e +2.319\times10^{25}\lambda_{\gamma}\cdot{\rm eV}\nonumber\\
	&-3.133\times10^{27}\lambda_g\cdot{\rm eV}-3.973\times10^{19}\lambda_d\nonumber\\
&	-3.967\times10^{19}\lambda_u \, .
\end{align}
According to Refs. \cite{PhysRevLett.120.141101,PhysRevLett.129.121102,Touboul2022}, the MICROSCOPE mission result (\ref{etaresult}) was achieved for a pair of test masses, which have different
compositions [PtRh(90/10) and TiAlV(90/6/4) alloys]. The scalar-charges for them are
\begin{align}
	&\zeta_{\rm Pt}=-1.496\times10^{18}\lambda_e+5.766\times10^{25}\lambda_{\gamma}\cdot{\rm eV}\nonumber\\
	&-3.149\times10^{27}\lambda_g\cdot{\rm eV}-4.320\times10^{19}\lambda_d\nonumber\\
	&-4.231\times10^{19}\lambda_u\label{alphaPt1} \\ 
	&\zeta_{\rm Ti}=-1.707\times10^{18}\lambda_e+3.101\times10^{25}\lambda_{\gamma}\cdot{\rm eV}\nonumber\\
	&-3.159\times10^{27}\lambda_g\cdot{\rm eV} -4.159\times10^{19}\lambda_d\nonumber\\
	&-4.123\times10^{19}\lambda_u \label{alphaTi1} 
\end{align}

Finally, it is clear that Eq. (\ref{etaequation1}) depends on five coupling parameters ($\lambda_e$,$\lambda_\gamma$,$\lambda_u$,$\lambda_d$, and $\lambda_g$), which is very different to the case of Eq. (\ref{deltaa1}), where $\delta a_l$ depends only on three coupling parameters ($\lambda_e$, $\lambda_\mu$ and $\lambda_\gamma$). 
To fully utilize these three experimental results ((\ref{daelectron}), (\ref{damuon}) and (\ref{etaresult})), let us mainly focus on the scalar-photon coupling $\lambda_\gamma$ and the scalar-lepton couplings ($\lambda_e$ and $\lambda_\mu$) in this work. In other words, we restrict our attention to the subspace, determined by $\lambda_u=\lambda_d=\lambda_g=0$, of the full six-dimensional parameter space. Then, the general linear coupling model (\ref{scalarmodel1}) becomes the reduced linear coupling model (\ref{scalarmodel}). Eqs.(\ref{earthcharge1}-\ref{alphaTi1}) are reduced to
\begin{align}
&\zeta'_{\rm E}=-1.808\times10^{18}\lambda_e +2.319\times10^{25}\lambda_{\gamma}\cdot{\rm eV}\,\,\,\,\,\,\nonumber\\
&\zeta'_{\rm Pt}=-1.496\times10^{18}\lambda_e+5.766\times10^{25}\lambda_{\gamma}\cdot{\rm eV}\nonumber\\
&\zeta'_{\rm Ti}=-1.707\times10^{18}\lambda_e +3.101\times10^{25}\lambda_{\gamma}\cdot{\rm eV} 
\end{align}

\bibliography{ScalarRefer2024.bib}

\begin{thebibliography}{50}%
\makeatletter
\providecommand \@ifxundefined [1]{%
 \@ifx{#1\undefined}
}%
\providecommand \@ifnum [1]{%
 \ifnum #1\expandafter \@firstoftwo
 \else \expandafter \@secondoftwo
 \fi
}%
\providecommand \@ifx [1]{%
 \ifx #1\expandafter \@firstoftwo
 \else \expandafter \@secondoftwo
 \fi
}%
\providecommand \natexlab [1]{#1}%
\providecommand \enquote  [1]{``#1''}%
\providecommand \bibnamefont  [1]{#1}%
\providecommand \bibfnamefont [1]{#1}%
\providecommand \citenamefont [1]{#1}%
\providecommand \href@noop [0]{\@secondoftwo}%
\providecommand \href [0]{\begingroup \@sanitize@url \@href}%
\providecommand \@href[1]{\@@startlink{#1}\@@href}%
\providecommand \@@href[1]{\endgroup#1\@@endlink}%
\providecommand \@sanitize@url [0]{\catcode `\\12\catcode `\$12\catcode
  `\&12\catcode `\#12\catcode `\^12\catcode `\_12\catcode `\%12\relax}%
\providecommand \@@startlink[1]{}%
\providecommand \@@endlink[0]{}%
\providecommand \url  [0]{\begingroup\@sanitize@url \@url }%
\providecommand \@url [1]{\endgroup\@href {#1}{\urlprefix }}%
\providecommand \urlprefix  [0]{URL }%
\providecommand \Eprint [0]{\href }%
\providecommand \doibase [0]{https://doi.org/}%
\providecommand \selectlanguage [0]{\@gobble}%
\providecommand \bibinfo  [0]{\@secondoftwo}%
\providecommand \bibfield  [0]{\@secondoftwo}%
\providecommand \translation [1]{[#1]}%
\providecommand \BibitemOpen [0]{}%
\providecommand \bibitemStop [0]{}%
\providecommand \bibitemNoStop [0]{.\EOS\space}%
\providecommand \EOS [0]{\spacefactor3000\relax}%
\providecommand \BibitemShut  [1]{\csname bibitem#1\endcsname}%
\let\auto@bib@innerbib\@empty
\bibitem [{\citenamefont {Peccei}\ and\ \citenamefont
  {Quinn}(1977)}]{PhysRevLett.38.1440}%
  \BibitemOpen
  \bibfield  {author} {\bibinfo {author} {\bibfnamefont {R.~D.}\ \bibnamefont
  {Peccei}}\ and\ \bibinfo {author} {\bibfnamefont {H.~R.}\ \bibnamefont
  {Quinn}},\ }\href {https://doi.org/10.1103/PhysRevLett.38.1440} {\bibfield
  {journal} {\bibinfo  {journal} {Phys. Rev. Lett.}\ }\textbf {\bibinfo
  {volume} {38}},\ \bibinfo {pages} {1440} (\bibinfo {year}
  {1977})}\BibitemShut {NoStop}%
\bibitem [{\citenamefont {Weinberg}(1978)}]{PhysRevLett.40.223}%
  \BibitemOpen
  \bibfield  {author} {\bibinfo {author} {\bibfnamefont {S.}~\bibnamefont
  {Weinberg}},\ }\href {https://doi.org/10.1103/PhysRevLett.40.223} {\bibfield
  {journal} {\bibinfo  {journal} {Phys. Rev. Lett.}\ }\textbf {\bibinfo
  {volume} {40}},\ \bibinfo {pages} {223} (\bibinfo {year} {1978})}\BibitemShut
  {NoStop}%
\bibitem [{\citenamefont {Wilczek}(1978)}]{PhysRevLett.40.279}%
  \BibitemOpen
  \bibfield  {author} {\bibinfo {author} {\bibfnamefont {F.}~\bibnamefont
  {Wilczek}},\ }\href {https://doi.org/10.1103/PhysRevLett.40.279} {\bibfield
  {journal} {\bibinfo  {journal} {Phys. Rev. Lett.}\ }\textbf {\bibinfo
  {volume} {40}},\ \bibinfo {pages} {279} (\bibinfo {year} {1978})}\BibitemShut
  {NoStop}%
\bibitem [{\citenamefont {Essig~R.}()}]{essig2013}%
  \BibitemOpen
  \bibfield  {author} {\bibinfo {author} {\bibfnamefont {{\it et
  al.}.}~\bibnamefont {Essig~R.}},\ }\bibfield  {journal} {\bibinfo  {journal}
  {arXiv: 1311.0029}\ }\href {https://doi.org/10.48550/arXiv.1311.0029}
  {10.48550/arXiv.1311.0029},\ \bibinfo {note} {, and references
  therein}\BibitemShut {NoStop}%
\bibitem [{\citenamefont {Patt}\ and\ \citenamefont
  {Wilczek}(2006)}]{Patt2006fw}%
  \BibitemOpen
  \bibfield  {author} {\bibinfo {author} {\bibfnamefont {B.}~\bibnamefont
  {Patt}}\ and\ \bibinfo {author} {\bibfnamefont {F.}~\bibnamefont {Wilczek}},\
  }\href@noop {} {\  (\bibinfo {year} {2006})},\ \Eprint
  {https://arxiv.org/abs/hep-ph/0605188} {arXiv:hep-ph/0605188} \BibitemShut
  {NoStop}%
\bibitem [{\citenamefont {O'Connell}\ \emph {et~al.}(2007)\citenamefont
  {O'Connell}, \citenamefont {Ramsey-Musolf},\ and\ \citenamefont
  {Wise}}]{PhysRevD.75.037701}%
  \BibitemOpen
  \bibfield  {author} {\bibinfo {author} {\bibfnamefont {D.}~\bibnamefont
  {O'Connell}}, \bibinfo {author} {\bibfnamefont {M.~J.}\ \bibnamefont
  {Ramsey-Musolf}},\ and\ \bibinfo {author} {\bibfnamefont {M.~B.}\
  \bibnamefont {Wise}},\ }\href {https://doi.org/10.1103/PhysRevD.75.037701}
  {\bibfield  {journal} {\bibinfo  {journal} {Phys. Rev. D}\ }\textbf {\bibinfo
  {volume} {75}},\ \bibinfo {pages} {037701} (\bibinfo {year}
  {2007})}\BibitemShut {NoStop}%
\bibitem [{\citenamefont {Barger}\ \emph {et~al.}(2008)\citenamefont {Barger},
  \citenamefont {Langacker}, \citenamefont {McCaskey}, \citenamefont
  {Ramsey-Musolf},\ and\ \citenamefont {Shaughnessy}}]{PhysRevD.77.035005}%
  \BibitemOpen
  \bibfield  {author} {\bibinfo {author} {\bibfnamefont {V.}~\bibnamefont
  {Barger}}, \bibinfo {author} {\bibfnamefont {P.}~\bibnamefont {Langacker}},
  \bibinfo {author} {\bibfnamefont {M.}~\bibnamefont {McCaskey}}, \bibinfo
  {author} {\bibfnamefont {M.~J.}\ \bibnamefont {Ramsey-Musolf}},\ and\
  \bibinfo {author} {\bibfnamefont {G.}~\bibnamefont {Shaughnessy}},\ }\href
  {https://doi.org/10.1103/PhysRevD.77.035005} {\bibfield  {journal} {\bibinfo
  {journal} {Phys. Rev. D}\ }\textbf {\bibinfo {volume} {77}},\ \bibinfo
  {pages} {035005} (\bibinfo {year} {2008})}\BibitemShut {NoStop}%
\bibitem [{\citenamefont {Piazza}\ and\ \citenamefont
  {Pospelov}(2010)}]{PhysRevD.82.043533}%
  \BibitemOpen
  \bibfield  {author} {\bibinfo {author} {\bibfnamefont {F.}~\bibnamefont
  {Piazza}}\ and\ \bibinfo {author} {\bibfnamefont {M.}~\bibnamefont
  {Pospelov}},\ }\href {https://doi.org/10.1103/PhysRevD.82.043533} {\bibfield
  {journal} {\bibinfo  {journal} {Phys. Rev. D}\ }\textbf {\bibinfo {volume}
  {82}},\ \bibinfo {pages} {043533} (\bibinfo {year} {2010})}\BibitemShut
  {NoStop}%
\bibitem [{\citenamefont {Conlon}(2006)}]{Conlon_2006}%
  \BibitemOpen
  \bibfield  {author} {\bibinfo {author} {\bibfnamefont {J.~P.}\ \bibnamefont
  {Conlon}},\ }\href {https://doi.org/10.1088/1126-6708/2006/05/078} {\bibfield
   {journal} {\bibinfo  {journal} {J. High Energy Phys.}\ }\textbf {\bibinfo
  {volume} {2006}}\bibinfo  {number} { (05)},\ \bibinfo {pages}
  {078}}\BibitemShut {NoStop}%
\bibitem [{\citenamefont {Svrcek}\ and\ \citenamefont
  {Witten}(2006)}]{Svrcek_2006}%
  \BibitemOpen
\bibfield  {number} {  }\bibfield  {author} {\bibinfo {author} {\bibfnamefont
  {P.}~\bibnamefont {Svrcek}}\ and\ \bibinfo {author} {\bibfnamefont
  {E.}~\bibnamefont {Witten}},\ }\href
  {https://doi.org/10.1088/1126-6708/2006/06/051} {\bibfield  {journal}
  {\bibinfo  {journal} {J. High Energy Phys.}\ }\textbf {\bibinfo {volume}
  {2006}}\bibinfo  {number} { (06)},\ \bibinfo {pages} {051}}\BibitemShut
  {NoStop}%
\bibitem [{\citenamefont {Arvanitaki}\ \emph {et~al.}(2010)\citenamefont
  {Arvanitaki}, \citenamefont {Dimopoulos}, \citenamefont {Dubovsky},
  \citenamefont {Kaloper},\ and\ \citenamefont
  {March-Russell}}]{PhysRevD.81.123530}%
  \BibitemOpen
\bibfield  {number} {  }\bibfield  {author} {\bibinfo {author} {\bibfnamefont
  {A.}~\bibnamefont {Arvanitaki}}, \bibinfo {author} {\bibfnamefont
  {S.}~\bibnamefont {Dimopoulos}}, \bibinfo {author} {\bibfnamefont
  {S.}~\bibnamefont {Dubovsky}}, \bibinfo {author} {\bibfnamefont
  {N.}~\bibnamefont {Kaloper}},\ and\ \bibinfo {author} {\bibfnamefont
  {J.}~\bibnamefont {March-Russell}},\ }\href
  {https://doi.org/10.1103/PhysRevD.81.123530} {\bibfield  {journal} {\bibinfo
  {journal} {Phys. Rev. D}\ }\textbf {\bibinfo {volume} {81}},\ \bibinfo
  {pages} {123530} (\bibinfo {year} {2010})}\BibitemShut {NoStop}%
\bibitem [{\citenamefont {Brans}\ and\ \citenamefont
  {Dicke}(1961)}]{PhysRev.124.925}%
  \BibitemOpen
  \bibfield  {author} {\bibinfo {author} {\bibfnamefont {C.}~\bibnamefont
  {Brans}}\ and\ \bibinfo {author} {\bibfnamefont {R.~H.}\ \bibnamefont
  {Dicke}},\ }\href {https://doi.org/10.1103/PhysRev.124.925} {\bibfield
  {journal} {\bibinfo  {journal} {Phys. Rev.}\ }\textbf {\bibinfo {volume}
  {124}},\ \bibinfo {pages} {925} (\bibinfo {year} {1961})}\BibitemShut
  {NoStop}%
\bibitem [{\citenamefont {Damour}\ and\ \citenamefont
  {Donoghue}(2010)}]{PhysRevD.82.084033}%
  \BibitemOpen
  \bibfield  {author} {\bibinfo {author} {\bibfnamefont {T.}~\bibnamefont
  {Damour}}\ and\ \bibinfo {author} {\bibfnamefont {J.~F.}\ \bibnamefont
  {Donoghue}},\ }\href {https://doi.org/10.1103/PhysRevD.82.084033} {\bibfield
  {journal} {\bibinfo  {journal} {Phys. Rev. D}\ }\textbf {\bibinfo {volume}
  {82}},\ \bibinfo {pages} {084033} (\bibinfo {year} {2010})}\BibitemShut
  {NoStop}%
\bibitem [{\citenamefont {Battaglieri~M.}()}]{battaglieri2017}%
  \BibitemOpen
  \bibfield  {author} {\bibinfo {author} {\bibfnamefont {{\it et
  al.}.}~\bibnamefont {Battaglieri~M.}},\ }\bibfield  {journal} {\bibinfo
  {journal} {arXiv: 1707.04591}\ }\href
  {https://doi.org/10.48550/arXiv.1707.04591}
  {10.48550/arXiv.1707.04591}\BibitemShut {NoStop}%
\bibitem [{\citenamefont {Navas~S.}(2024)}]{PhysRevD.110.030001}%
  \BibitemOpen
  \bibfield  {author} {\bibinfo {author} {\bibfnamefont {{\it et
  al.}.}~\bibnamefont {Navas~S.}} (\bibinfo {collaboration} {Particle Data
  Group Collaboration}),\ }\href {https://doi.org/10.1103/PhysRevD.110.030001}
  {\bibfield  {journal} {\bibinfo  {journal} {Phys. Rev. D}\ }\textbf {\bibinfo
  {volume} {110}},\ \bibinfo {pages} {030001} (\bibinfo {year} {2024})},\
  \bibinfo {note} {\,Section 90}\BibitemShut {NoStop}%
\bibitem [{\citenamefont {Aoyama}\ \emph
  {et~al.}(2012{\natexlab{a}})\citenamefont {Aoyama}, \citenamefont {Hayakawa},
  \citenamefont {Kinoshita},\ and\ \citenamefont
  {Nio}}]{PhysRevLett.109.111807}%
  \BibitemOpen
  \bibfield  {author} {\bibinfo {author} {\bibfnamefont {T.}~\bibnamefont
  {Aoyama}}, \bibinfo {author} {\bibfnamefont {M.}~\bibnamefont {Hayakawa}},
  \bibinfo {author} {\bibfnamefont {T.}~\bibnamefont {Kinoshita}},\ and\
  \bibinfo {author} {\bibfnamefont {M.}~\bibnamefont {Nio}},\ }\href
  {https://doi.org/10.1103/PhysRevLett.109.111807} {\bibfield  {journal}
  {\bibinfo  {journal} {Phys. Rev. Lett.}\ }\textbf {\bibinfo {volume} {109}},\
  \bibinfo {pages} {111807} (\bibinfo {year} {2012}{\natexlab{a}})}\BibitemShut
  {NoStop}%
\bibitem [{\citenamefont {Aoyama}\ \emph {et~al.}(2019)\citenamefont {Aoyama},
  \citenamefont {Kinoshita},\ and\ \citenamefont {Nio}}]{atoms7010028}%
  \BibitemOpen
  \bibfield  {author} {\bibinfo {author} {\bibfnamefont {T.}~\bibnamefont
  {Aoyama}}, \bibinfo {author} {\bibfnamefont {T.}~\bibnamefont {Kinoshita}},\
  and\ \bibinfo {author} {\bibfnamefont {M.}~\bibnamefont {Nio}},\ }\href
  {https://doi.org/10.3390/atoms7010028} {\bibfield  {journal} {\bibinfo
  {journal} {Atoms}\ }\textbf {\bibinfo {volume} {7}},\ \bibinfo {pages} {28}
  (\bibinfo {year} {2019})}\BibitemShut {NoStop}%
\bibitem [{\citenamefont {Volkov}()}]{volkov2024}%
  \BibitemOpen
  \bibfield  {author} {\bibinfo {author} {\bibfnamefont {S.}~\bibnamefont
  {Volkov}},\ }\bibfield  {journal} {\bibinfo  {journal} {arXiv:2404.00649}\
  }\href {https://doi.org/10.48550/arXiv.2404.00649}
  {10.48550/arXiv.2404.00649}\BibitemShut {NoStop}%
\bibitem [{\citenamefont {Aoyama}\ \emph
  {et~al.}(2012{\natexlab{b}})\citenamefont {Aoyama}, \citenamefont {Hayakawa},
  \citenamefont {Kinoshita},\ and\ \citenamefont
  {Nio}}]{PhysRevLett.109.111808}%
  \BibitemOpen
  \bibfield  {author} {\bibinfo {author} {\bibfnamefont {T.}~\bibnamefont
  {Aoyama}}, \bibinfo {author} {\bibfnamefont {M.}~\bibnamefont {Hayakawa}},
  \bibinfo {author} {\bibfnamefont {T.}~\bibnamefont {Kinoshita}},\ and\
  \bibinfo {author} {\bibfnamefont {M.}~\bibnamefont {Nio}},\ }\href
  {https://doi.org/10.1103/PhysRevLett.109.111808} {\bibfield  {journal}
  {\bibinfo  {journal} {Phys. Rev. Lett.}\ }\textbf {\bibinfo {volume} {109}},\
  \bibinfo {pages} {111808} (\bibinfo {year} {2012}{\natexlab{b}})}\BibitemShut
  {NoStop}%
\bibitem [{\citenamefont {Aoyama~T.}(2020)}]{AOYAMA20201}%
  \BibitemOpen
  \bibfield  {author} {\bibinfo {author} {\bibfnamefont {{\it et
  al.}.}~\bibnamefont {Aoyama~T.}},\ }\href
  {https://doi.org/https://doi.org/10.1016/j.physrep.2020.07.006} {\bibfield
  {journal} {\bibinfo  {journal} {Phys. Rept.}\ }\textbf {\bibinfo {volume}
  {887}},\ \bibinfo {pages} {1} (\bibinfo {year} {2020})}\BibitemShut {NoStop}%
\bibitem [{\citenamefont {Colangelo~G.}()}]{colangelo2022prospects}%
  \BibitemOpen
  \bibfield  {author} {\bibinfo {author} {\bibfnamefont {{\it et
  al.}.}~\bibnamefont {Colangelo~G.}},\ }\bibfield  {journal} {\bibinfo
  {journal} {arXiv: 2203.15810}\ }\href
  {https://doi.org/10.48550/arXiv.2203.15810}
  {10.48550/arXiv.2203.15810}\BibitemShut {NoStop}%
\bibitem [{\citenamefont {Parker}\ \emph {et~al.}(2018)\citenamefont {Parker},
  \citenamefont {Yu}, \citenamefont {Zhong}, \citenamefont {Estey},\ and\
  \citenamefont {Müller}}]{Parker2018}%
  \BibitemOpen
  \bibfield  {author} {\bibinfo {author} {\bibfnamefont {R.~H.}\ \bibnamefont
  {Parker}}, \bibinfo {author} {\bibfnamefont {C.}~\bibnamefont {Yu}}, \bibinfo
  {author} {\bibfnamefont {W.}~\bibnamefont {Zhong}}, \bibinfo {author}
  {\bibfnamefont {B.}~\bibnamefont {Estey}},\ and\ \bibinfo {author}
  {\bibfnamefont {H.}~\bibnamefont {Müller}},\ }\href
  {https://doi.org/10.1126/science.aap7706} {\bibfield  {journal} {\bibinfo
  {journal} {Science}\ }\textbf {\bibinfo {volume} {360}},\ \bibinfo {pages}
  {191} (\bibinfo {year} {2018})}\BibitemShut {NoStop}%
\bibitem [{\citenamefont {Morel}\ \emph {et~al.}(2020)\citenamefont {Morel},
  \citenamefont {Yao}, \citenamefont {Cladé},\ and\ \citenamefont
  {Guellati-Khélifa}}]{Morel2020}%
  \BibitemOpen
  \bibfield  {author} {\bibinfo {author} {\bibfnamefont {L.}~\bibnamefont
  {Morel}}, \bibinfo {author} {\bibfnamefont {Z.}~\bibnamefont {Yao}}, \bibinfo
  {author} {\bibfnamefont {P.}~\bibnamefont {Cladé}},\ and\ \bibinfo {author}
  {\bibfnamefont {S.}~\bibnamefont {Guellati-Khélifa}},\ }\href
  {https://doi.org/10.1038/s41586-020-2964-7} {\bibfield  {journal} {\bibinfo
  {journal} {Nature}\ }\textbf {\bibinfo {volume} {588}},\ \bibinfo {pages}
  {61} (\bibinfo {year} {2020})}\BibitemShut {NoStop}%
\bibitem [{\citenamefont {Chang}\ \emph {et~al.}(2001)\citenamefont {Chang},
  \citenamefont {Chang}, \citenamefont {Chou},\ and\ \citenamefont
  {Keung}}]{PhysRevD.63.091301}%
  \BibitemOpen
  \bibfield  {author} {\bibinfo {author} {\bibfnamefont {D.}~\bibnamefont
  {Chang}}, \bibinfo {author} {\bibfnamefont {W.-F.}\ \bibnamefont {Chang}},
  \bibinfo {author} {\bibfnamefont {C.-H.}\ \bibnamefont {Chou}},\ and\
  \bibinfo {author} {\bibfnamefont {W.-Y.}\ \bibnamefont {Keung}},\ }\href
  {https://doi.org/10.1103/PhysRevD.63.091301} {\bibfield  {journal} {\bibinfo
  {journal} {Phys. Rev. D}\ }\textbf {\bibinfo {volume} {63}},\ \bibinfo
  {pages} {091301} (\bibinfo {year} {2001})}\BibitemShut {NoStop}%
\bibitem [{\citenamefont {Chen}\ \emph {et~al.}(2016)\citenamefont {Chen},
  \citenamefont {Davoudiasl}, \citenamefont {Marciano},\ and\ \citenamefont
  {Zhang}}]{PhysRevD.93.035006}%
  \BibitemOpen
  \bibfield  {author} {\bibinfo {author} {\bibfnamefont {C.-Y.}\ \bibnamefont
  {Chen}}, \bibinfo {author} {\bibfnamefont {H.}~\bibnamefont {Davoudiasl}},
  \bibinfo {author} {\bibfnamefont {W.~J.}\ \bibnamefont {Marciano}},\ and\
  \bibinfo {author} {\bibfnamefont {C.}~\bibnamefont {Zhang}},\ }\href
  {https://doi.org/10.1103/PhysRevD.93.035006} {\bibfield  {journal} {\bibinfo
  {journal} {Phys. Rev. D}\ }\textbf {\bibinfo {volume} {93}},\ \bibinfo
  {pages} {035006} (\bibinfo {year} {2016})}\BibitemShut {NoStop}%
\bibitem [{\citenamefont {Marciano}\ \emph {et~al.}(2016)\citenamefont
  {Marciano}, \citenamefont {Masiero}, \citenamefont {Paradisi},\ and\
  \citenamefont {Passera}}]{PhysRevD.94.115033}%
  \BibitemOpen
  \bibfield  {author} {\bibinfo {author} {\bibfnamefont {W.~J.}\ \bibnamefont
  {Marciano}}, \bibinfo {author} {\bibfnamefont {A.}~\bibnamefont {Masiero}},
  \bibinfo {author} {\bibfnamefont {P.}~\bibnamefont {Paradisi}},\ and\
  \bibinfo {author} {\bibfnamefont {M.}~\bibnamefont {Passera}},\ }\href
  {https://doi.org/10.1103/PhysRevD.94.115033} {\bibfield  {journal} {\bibinfo
  {journal} {Phys. Rev. D}\ }\textbf {\bibinfo {volume} {94}},\ \bibinfo
  {pages} {115033} (\bibinfo {year} {2016})}\BibitemShut {NoStop}%
\bibitem [{\citenamefont {Fan}\ \emph {et~al.}(2023)\citenamefont {Fan},
  \citenamefont {Myers}, \citenamefont {Sukra},\ and\ \citenamefont
  {Gabrielse}}]{Fan2023}%
  \BibitemOpen
  \bibfield  {author} {\bibinfo {author} {\bibfnamefont {X.}~\bibnamefont
  {Fan}}, \bibinfo {author} {\bibfnamefont {T.~G.}\ \bibnamefont {Myers}},
  \bibinfo {author} {\bibfnamefont {B.~A.~D.}\ \bibnamefont {Sukra}},\ and\
  \bibinfo {author} {\bibfnamefont {G.}~\bibnamefont {Gabrielse}},\ }\href
  {https://doi.org/10.1103/PhysRevLett.130.071801} {\bibfield  {journal}
  {\bibinfo  {journal} {Phys. Rev. Lett.}\ }\textbf {\bibinfo {volume} {130}},\
  \bibinfo {pages} {071801} (\bibinfo {year} {2023})}\BibitemShut {NoStop}%
\bibitem [{\citenamefont {Aguillard D.~P.}(2023)}]{PhysRevLett.131.161802}%
  \BibitemOpen
  \bibfield  {author} {\bibinfo {author} {\bibfnamefont {{\it et
  al.}.}~\bibnamefont {Aguillard D.~P.}} (\bibinfo {collaboration} {The Muon
  $g\ensuremath{-}2$ Collaboration}),\ }\href
  {https://doi.org/10.1103/PhysRevLett.131.161802} {\bibfield  {journal}
  {\bibinfo  {journal} {Phys. Rev. Lett.}\ }\textbf {\bibinfo {volume} {131}},\
  \bibinfo {pages} {161802} (\bibinfo {year} {2023})}\BibitemShut {NoStop}%
\bibitem [{\citenamefont {Aliberti~R.}()}]{wp25}%
  \BibitemOpen
  \bibfield  {author} {\bibinfo {author} {\bibfnamefont {{\it et
  al.}.}~\bibnamefont {Aliberti~R.}},\ }\bibfield  {journal} {\bibinfo
  {journal} {arXiv: 2505.21476}\ }\href
  {https://doi.org/10.48550/arXiv.2505.21476} {10.48550/arXiv.2505.21476},\
  \bibinfo {note} {, and references therein}\BibitemShut {NoStop}%
\bibitem [{\citenamefont {Aguillard D.~P.}()}]{muon2025}%
  \BibitemOpen
  \bibfield  {author} {\bibinfo {author} {\bibfnamefont {{\it et
  al.}.}~\bibnamefont {Aguillard D.~P.}},\ }\bibfield  {journal} {\bibinfo
  {journal} {arXiv: 2506.03069}\ }\href
  {https://doi.org/10.48550/arXiv.2506.03069} {10.48550/arXiv.2506.03069},\
  \bibinfo {note} {, and references therein}\BibitemShut {NoStop}%
\bibitem [{\citenamefont {Berg\'e}\ \emph {et~al.}(2018)\citenamefont
  {Berg\'e}, \citenamefont {Brax}, \citenamefont {M\'etris}, \citenamefont
  {Pernot-Borr\`as}, \citenamefont {Touboul},\ and\ \citenamefont
  {Uzan}}]{PhysRevLett.120.141101}%
  \BibitemOpen
  \bibfield  {author} {\bibinfo {author} {\bibfnamefont {J.}~\bibnamefont
  {Berg\'e}}, \bibinfo {author} {\bibfnamefont {P.}~\bibnamefont {Brax}},
  \bibinfo {author} {\bibfnamefont {G.}~\bibnamefont {M\'etris}}, \bibinfo
  {author} {\bibfnamefont {M.}~\bibnamefont {Pernot-Borr\`as}}, \bibinfo
  {author} {\bibfnamefont {P.}~\bibnamefont {Touboul}},\ and\ \bibinfo {author}
  {\bibfnamefont {J.-P.}\ \bibnamefont {Uzan}},\ }\href
  {https://doi.org/10.1103/PhysRevLett.120.141101} {\bibfield  {journal}
  {\bibinfo  {journal} {Phys. Rev. Lett.}\ }\textbf {\bibinfo {volume} {120}},\
  \bibinfo {pages} {141101} (\bibinfo {year} {2018})}\BibitemShut {NoStop}%
\bibitem [{\citenamefont
  {Touboul~P.}(2022{\natexlab{a}})}]{PhysRevLett.129.121102}%
  \BibitemOpen
  \bibfield  {author} {\bibinfo {author} {\bibfnamefont {{\it et
  al.}.}~\bibnamefont {Touboul~P.}} (\bibinfo {collaboration} {MICROSCOPE
  Collaboration}),\ }\href {https://doi.org/10.1103/PhysRevLett.129.121102}
  {\bibfield  {journal} {\bibinfo  {journal} {Phys. Rev. Lett.}\ }\textbf
  {\bibinfo {volume} {129}},\ \bibinfo {pages} {121102} (\bibinfo {year}
  {2022}{\natexlab{a}})}\BibitemShut {NoStop}%
\bibitem [{\citenamefont {Touboul~P.}(2022{\natexlab{b}})}]{Touboul2022}%
  \BibitemOpen
  \bibfield  {author} {\bibinfo {author} {\bibfnamefont {{\it et
  al.}.}~\bibnamefont {Touboul~P.}} (\bibinfo {collaboration} {MICROSCOPE
  Collaboration}),\ }\href {https://doi.org/10.1088/1361-6382/ac84be}
  {\bibfield  {journal} {\bibinfo  {journal} {Class. Quant. Grav.}\ }\textbf
  {\bibinfo {volume} {39}},\ \bibinfo {pages} {204009} (\bibinfo {year}
  {2022}{\natexlab{b}})}\BibitemShut {NoStop}%
\bibitem [{\citenamefont {Bergé}(2023)}]{Berge_2023}%
  \BibitemOpen
  \bibfield  {author} {\bibinfo {author} {\bibfnamefont {J.}~\bibnamefont
  {Bergé}},\ }\href {https://doi.org/10.1088/1361-6633/acd203} {\bibfield
  {journal} {\bibinfo  {journal} {Rept. Prog. Phys.}\ }\textbf {\bibinfo
  {volume} {86}},\ \bibinfo {pages} {066901} (\bibinfo {year}
  {2023})}\BibitemShut {NoStop}%
\bibitem [{\citenamefont {Beloy~K.}(2021)}]{Beloy2021}%
  \BibitemOpen
  \bibfield  {author} {\bibinfo {author} {\bibfnamefont {{\it et
  al.}.}~\bibnamefont {Beloy~K.}} (\bibinfo {collaboration} {Boulder Atomic
  Clock Optical Network (BACON)}),\ }\href
  {https://doi.org/10.1038/s41586-021-03253-4} {\bibfield  {journal} {\bibinfo
  {journal} {Nature}\ }\textbf {\bibinfo {volume} {591}},\ \bibinfo {pages}
  {564} (\bibinfo {year} {2021})}\BibitemShut {NoStop}%
\bibitem [{\citenamefont {Vermeulen S.~M.}(2021)}]{Vermeulen2021}%
  \BibitemOpen
  \bibfield  {author} {\bibinfo {author} {\bibfnamefont {{\it et
  al.}.}~\bibnamefont {Vermeulen S.~M.}},\ }\href
  {https://doi.org/10.1038/s41586-021-04031-y} {\bibfield  {journal} {\bibinfo
  {journal} {Nature}\ }\textbf {\bibinfo {volume} {600}},\ \bibinfo {pages}
  {424} (\bibinfo {year} {2021})}\BibitemShut {NoStop}%
\bibitem [{\citenamefont {Passarino}\ and\ \citenamefont
  {Veltman}(1979)}]{PASSARINO1979151}%
  \BibitemOpen
  \bibfield  {author} {\bibinfo {author} {\bibfnamefont {G.}~\bibnamefont
  {Passarino}}\ and\ \bibinfo {author} {\bibfnamefont {M.}~\bibnamefont
  {Veltman}},\ }\href
  {https://doi.org/https://doi.org/10.1016/0550-3213(79)90234-7} {\bibfield
  {journal} {\bibinfo  {journal} {Nucl. Phys. B}\ }\textbf {\bibinfo {volume}
  {160}},\ \bibinfo {pages} {151} (\bibinfo {year} {1979})}\BibitemShut
  {NoStop}%
\bibitem [{\citenamefont {{'t Hooft}}\ and\ \citenamefont
  {Veltman}(1979)}]{THOOFT1979365}%
  \BibitemOpen
  \bibfield  {author} {\bibinfo {author} {\bibfnamefont {G.}~\bibnamefont {{'t
  Hooft}}}\ and\ \bibinfo {author} {\bibfnamefont {M.}~\bibnamefont
  {Veltman}},\ }\href
  {https://doi.org/https://doi.org/10.1016/0550-3213(79)90605-9} {\bibfield
  {journal} {\bibinfo  {journal} {Nucl. Phys. B}\ }\textbf {\bibinfo {volume}
  {153}},\ \bibinfo {pages} {365} (\bibinfo {year} {1979})}\BibitemShut
  {NoStop}%
\bibitem [{\citenamefont {Zhao}\ \emph {et~al.}(2022)\citenamefont {Zhao},
  \citenamefont {Gao}, \citenamefont {Wang},\ and\ \citenamefont
  {Zhan}}]{Zhao2022}%
  \BibitemOpen
  \bibfield  {author} {\bibinfo {author} {\bibfnamefont {W.}~\bibnamefont
  {Zhao}}, \bibinfo {author} {\bibfnamefont {D.}~\bibnamefont {Gao}}, \bibinfo
  {author} {\bibfnamefont {J.}~\bibnamefont {Wang}},\ and\ \bibinfo {author}
  {\bibfnamefont {M.}~\bibnamefont {Zhan}},\ }\href
  {https://doi.org/10.1007/s10714-022-02925-4} {\bibfield  {journal} {\bibinfo
  {journal} {Gen. Relat. Grav.}\ }\textbf {\bibinfo {volume} {54}},\ \bibinfo
  {pages} {41} (\bibinfo {year} {2022})}\BibitemShut {NoStop}%
\bibitem [{\citenamefont {Hardy}\ and\ \citenamefont
  {Lasenby}(2017)}]{Hardy2017}%
  \BibitemOpen
  \bibfield  {author} {\bibinfo {author} {\bibfnamefont {E.}~\bibnamefont
  {Hardy}}\ and\ \bibinfo {author} {\bibfnamefont {R.}~\bibnamefont
  {Lasenby}},\ }\href {https://doi.org/10.1007/JHEP02(2017)033} {\bibfield
  {journal} {\bibinfo  {journal} {J. High Energy Phys.}\ }\textbf {\bibinfo
  {volume} {2017}}\bibinfo  {number} { (2)},\ \bibinfo {pages}
  {33}}\BibitemShut {NoStop}%
\bibitem [{\citenamefont {Knapen}\ \emph {et~al.}(2017)\citenamefont {Knapen},
  \citenamefont {Lin},\ and\ \citenamefont {Zurek}}]{PhysRevD.96.115021}%
  \BibitemOpen
\bibfield  {number} {  }\bibfield  {author} {\bibinfo {author} {\bibfnamefont
  {S.}~\bibnamefont {Knapen}}, \bibinfo {author} {\bibfnamefont
  {T.}~\bibnamefont {Lin}},\ and\ \bibinfo {author} {\bibfnamefont {K.~M.}\
  \bibnamefont {Zurek}},\ }\href {https://doi.org/10.1103/PhysRevD.96.115021}
  {\bibfield  {journal} {\bibinfo  {journal} {Phys. Rev. D}\ }\textbf {\bibinfo
  {volume} {96}},\ \bibinfo {pages} {115021} (\bibinfo {year}
  {2017})}\BibitemShut {NoStop}%
\bibitem [{\citenamefont {Giudice}\ \emph {et~al.}(2012)\citenamefont
  {Giudice}, \citenamefont {Paradisi},\ and\ \citenamefont
  {Passera}}]{Giudice2012}%
  \BibitemOpen
  \bibfield  {author} {\bibinfo {author} {\bibfnamefont {G.~F.}\ \bibnamefont
  {Giudice}}, \bibinfo {author} {\bibfnamefont {P.}~\bibnamefont {Paradisi}},\
  and\ \bibinfo {author} {\bibfnamefont {M.}~\bibnamefont {Passera}},\ }\href
  {https://doi.org/10.1007/JHEP11(2012)113} {\bibfield  {journal} {\bibinfo
  {journal} {J. High Energy Phys.}\ }\textbf {\bibinfo {volume} {2012}}\bibinfo
   {number} { (11)},\ \bibinfo {pages} {113}}\BibitemShut {NoStop}%
\bibitem [{\citenamefont {Struckmann}\ \emph {et~al.}(2024)\citenamefont
  {Struckmann}, \citenamefont {Corgier}, \citenamefont {Loriani}, \citenamefont
  {Kleinsteinberg}, \citenamefont {Gox}, \citenamefont {Giese}, \citenamefont
  {M\'etris}, \citenamefont {Gaaloul},\ and\ \citenamefont
  {Wolf}}]{PhysRevD.109.064010}%
  \BibitemOpen
\bibfield  {number} {  }\bibfield  {author} {\bibinfo {author} {\bibfnamefont
  {C.}~\bibnamefont {Struckmann}}, \bibinfo {author} {\bibfnamefont
  {R.}~\bibnamefont {Corgier}}, \bibinfo {author} {\bibfnamefont
  {S.}~\bibnamefont {Loriani}}, \bibinfo {author} {\bibfnamefont
  {G.}~\bibnamefont {Kleinsteinberg}}, \bibinfo {author} {\bibfnamefont
  {N.}~\bibnamefont {Gox}}, \bibinfo {author} {\bibfnamefont {E.}~\bibnamefont
  {Giese}}, \bibinfo {author} {\bibfnamefont {G.}~\bibnamefont {M\'etris}},
  \bibinfo {author} {\bibfnamefont {N.}~\bibnamefont {Gaaloul}},\ and\ \bibinfo
  {author} {\bibfnamefont {P.}~\bibnamefont {Wolf}},\ }\href
  {https://doi.org/10.1103/PhysRevD.109.064010} {\bibfield  {journal} {\bibinfo
   {journal} {Phys. Rev. D}\ }\textbf {\bibinfo {volume} {109}},\ \bibinfo
  {pages} {064010} (\bibinfo {year} {2024})}\BibitemShut {NoStop}%
\bibitem [{\citenamefont {Nowakowski}\ \emph {et~al.}(2005)\citenamefont
  {Nowakowski}, \citenamefont {Paschos},\ and\ \citenamefont
  {Rodríguez}}]{Nowakowski2005}%
  \BibitemOpen
  \bibfield  {author} {\bibinfo {author} {\bibfnamefont {M.}~\bibnamefont
  {Nowakowski}}, \bibinfo {author} {\bibfnamefont {E.~A.}\ \bibnamefont
  {Paschos}},\ and\ \bibinfo {author} {\bibfnamefont {J.~M.}\ \bibnamefont
  {Rodríguez}},\ }\href {https://doi.org/10.1088/0143-0807/26/4/001}
  {\bibfield  {journal} {\bibinfo  {journal} {Eur. J. Phys.}\ }\textbf
  {\bibinfo {volume} {26}},\ \bibinfo {pages} {545} (\bibinfo {year}
  {2005})}\BibitemShut {NoStop}%
\bibitem [{\citenamefont {Mertig}\ \emph {et~al.}(1991)\citenamefont {Mertig},
  \citenamefont {Böhm},\ and\ \citenamefont {Denner}}]{MERTIG1991345}%
  \BibitemOpen
  \bibfield  {author} {\bibinfo {author} {\bibfnamefont {R.}~\bibnamefont
  {Mertig}}, \bibinfo {author} {\bibfnamefont {M.}~\bibnamefont {Böhm}},\ and\
  \bibinfo {author} {\bibfnamefont {A.}~\bibnamefont {Denner}},\ }\href
  {https://doi.org/https://doi.org/10.1016/0010-4655(91)90130-D} {\bibfield
  {journal} {\bibinfo  {journal} {Comp. Phys. Comm.}\ }\textbf {\bibinfo
  {volume} {64}},\ \bibinfo {pages} {345} (\bibinfo {year} {1991})}\BibitemShut
  {NoStop}%
\bibitem [{\citenamefont {Shtabovenko}\ \emph {et~al.}(2020)\citenamefont
  {Shtabovenko}, \citenamefont {Mertig},\ and\ \citenamefont
  {Orellana}}]{SHTABOVENKO2020107478}%
  \BibitemOpen
  \bibfield  {author} {\bibinfo {author} {\bibfnamefont {V.}~\bibnamefont
  {Shtabovenko}}, \bibinfo {author} {\bibfnamefont {R.}~\bibnamefont
  {Mertig}},\ and\ \bibinfo {author} {\bibfnamefont {F.}~\bibnamefont
  {Orellana}},\ }\href
  {https://doi.org/https://doi.org/10.1016/j.cpc.2020.107478} {\bibfield
  {journal} {\bibinfo  {journal} {Comp. Phys. Comm.}\ }\textbf {\bibinfo
  {volume} {256}},\ \bibinfo {pages} {107478} (\bibinfo {year}
  {2020})}\BibitemShut {NoStop}%
\bibitem [{\citenamefont {Hahn}(2001)}]{HAHN2001418}%
  \BibitemOpen
  \bibfield  {author} {\bibinfo {author} {\bibfnamefont {T.}~\bibnamefont
  {Hahn}},\ }\href
  {https://doi.org/https://doi.org/10.1016/S0010-4655(01)00290-9} {\bibfield
  {journal} {\bibinfo  {journal} {Comp. Phys. Comm.}\ }\textbf {\bibinfo
  {volume} {140}},\ \bibinfo {pages} {418} (\bibinfo {year}
  {2001})}\BibitemShut {NoStop}%
\bibitem [{\citenamefont {Patel}(2015)}]{PATEL2015276}%
  \BibitemOpen
  \bibfield  {author} {\bibinfo {author} {\bibfnamefont {H.~H.}\ \bibnamefont
  {Patel}},\ }\href {https://doi.org/https://doi.org/10.1016/j.cpc.2015.08.017}
  {\bibfield  {journal} {\bibinfo  {journal} {Comp. Phys. Comm.}\ }\textbf
  {\bibinfo {volume} {197}},\ \bibinfo {pages} {276} (\bibinfo {year}
  {2015})}\BibitemShut {NoStop}%
\bibitem [{\citenamefont {Patel}(2017)}]{PATEL201766}%
  \BibitemOpen
  \bibfield  {author} {\bibinfo {author} {\bibfnamefont {H.~H.}\ \bibnamefont
  {Patel}},\ }\href {https://doi.org/https://doi.org/10.1016/j.cpc.2017.04.015}
  {\bibfield  {journal} {\bibinfo  {journal} {Comp. Phys. Comm.}\ }\textbf
  {\bibinfo {volume} {218}},\ \bibinfo {pages} {66} (\bibinfo {year}
  {2017})}\BibitemShut {NoStop}%
\bibitem [{\citenamefont {Morgan}\ and\ \citenamefont
  {Anders}(1980)}]{pnas.77.12.6973}%
  \BibitemOpen
  \bibfield  {author} {\bibinfo {author} {\bibfnamefont {J.~W.}\ \bibnamefont
  {Morgan}}\ and\ \bibinfo {author} {\bibfnamefont {E.}~\bibnamefont
  {Anders}},\ }\href {https://doi.org/10.1073/pnas.77.12.6973} {\bibfield
  {journal} {\bibinfo  {journal} {Proc. Natl. Acad. Sci. USA}\ }\textbf
  {\bibinfo {volume} {77}},\ \bibinfo {pages} {6973} (\bibinfo {year}
  {1980})}\BibitemShut {NoStop}%
\end{thebibliography}%

\end{document}